\documentstyle[12pt]{article}
\def\be{\begin{equation}}
\def\ee{\end{equation}}
\def\bq{\begin{eqnarray}}
\def\eq{\end{eqnarray}}
\newcommand{\ra}{\rightarrow}
\newcommand{\g}{\gamma}

\newcommand{\vp}{\varphi}
\newcommand{\al}{\alpha}
\newcommand{\bet}{\beta}

\newcounter{saveeqn}
\newcounter{App} 
\newcommand{\app}{%
\stepcounter{App}%
\setcounter{saveeqn}{\value{equation}}%
\setcounter{equation}{0}%
\renewcommand{\theequation}{\Alph{App}\arabic{equation}} }
\newcommand{\appende}{%
\setcounter{equation}{\value{saveeqn}}%
\renewcommand{\theequation}{\arabic{equation}}  }

\begin{document}

\thispagestyle{empty}
\setcounter{page}{0}

\begin{flushright}
MPI-PhT/94-62 \\
CEBAF-TH-94-22\\
LMU 15/94\\
September 1994
\end{flushright}
\vspace*{\fill}
\begin{center}
{\Large \bf $D^*D\pi$ and $B^*B\pi$ couplings in QCD $^+$}\\
\vspace{2em}
\large
{\bf V.M. Belyaev$^{1,a }$,
V.M. Braun$^{2,b}$, A. Khodjamirian$^{3,c *}$,
R. R\"uckl$^{2,3}$}\\
\vspace{2em}

{$^1$ \small CEBAF, Newport News, Virginia 23606, USA} \\

{$^2$ \small Max-Planck-Institut f\"ur Physik, Werner-Heisenberg-Institut,
D-80805 M\"unchen, Germany }\\

{$^3$ \small Sektion Physik der Universit\"at M\"unchen, D-80333 M\"unchen ,
Germany }\\

\end{center}
\vspace*{\fill}

\begin{abstract}

We calculate the $D^*D\pi$ and $B^*B\pi$ couplings
using QCD sum rules on the light-cone. In this approach, the large-distance
dynamics is incorporated in a set of pion wave functions.
We take into account two-particle and three-particle wave functions of
twist 2, 3 and 4.
The resulting values of the coupling constants are
$g_{D^*D\pi}= 12.5\pm 1$ and $g_{B^*B\pi}= 29\pm 3 $. From this
we predict the partial width
$ \Gamma (D^{*+} \ra D^0 \pi^+ )=32 \pm 5~ keV $.
We also discuss the soft-pion limit
of the sum rules which is equivalent to the external
axial field approach employed in earlier calculations.
Furthermore, using  $g_{B^*B\pi}$ and
$g_{D^*D\pi}$ the pole dominance model for the $ B \ra \pi$ and
$D\ra \pi$ semileptonic form factors
is compared with the direct calculation of these
form factors in the same framework of light-cone sum rules.

\end{abstract}

\vspace*{\fill}

\begin{flushleft}
{$^a$ \small on leave from Institute of Theoretical and Experimental Physics,
117259 Moscow, Russia }\\
{$^b$ \small on leave from St.Petersburg Nuclear Physics Institute,
188350 Gatchina, Russia; now at DESY, D--22603 Hamburg, Germany} \\
{$^c$ \small on leave from Yerevan Physics Institute, 375036 Yerevan, Armenia }
\\
{$^{*}$\it Alexander von Humboldt Fellow } \\
\noindent$^+$ Work supported by the German Federal Ministry for Research and
Technology under contract No. 05 6MU93P\\

\end{flushleft}

\newpage

\section{Introduction}
The extraction of fundamental parameters from data on heavy
flavoured hadrons inevitably requires some information about the
physics at large distances. Numerous theoretical studies
have been devoted to making this extraction as reliable
as possible. While the inclusive $B$ and $D$ decays appear to
be the most clean reactions theoretically, exclusive decays are
often much easier to measure experimentally. However, for their
interpretation one needs accurate estimates of
decay form factors and other hadronic
matrix elements. In the exceptional case $B \ra D e \nu $,
the form factor at zero recoil can
be calculated in the heavy quark limit \cite{IW89,N91}. In most other
important cases, one has to rely on less rigorous nonperturbative approaches.
Among those, QCD sum rules \cite{SVZ} have proved
to be particularly powerful.

In this paper we employ sum rule methods in order to calculate
the $D^*D\pi$ and $B^*B\pi$ coupling constants. These couplings
are interesting for several reasons. In particular, they
determine the normalization of the
heavy-to-light form factors $D \ra \pi $ and
$B \ra \pi $ near zero pionic recoil, where the $D^*$- and
$B^*$- poles are believed to dominate. For further discussion one may
consult refs. \cite{IW90,BLN}. Recently, it
has been argued
that in the combined heavy quark and chiral limit vector meson
dominance becomes even exact \cite{GM}.
As noted in refs. \cite{BBD1,BKR}, $B^*$ dominance is also compatible
with the dependence of the $B \ra \pi$ form factor
on the momentum transfer $p$ predicted by QCD sum rules
at low values of $p^2$. A similar conclusion is reached in ref. \cite{Ball}
concerning $D^*$ dominance in the $D \ra \pi$ form factor.
In addition, the $D^*D\pi$ coupling can be directly measured in
the decay $D^* \ra D\pi$ and thus provides one more independent test
of the sum rule approach.

Calculations of couplings of heavy mesons to a pion
have already been undertaken several times in the framework of
QCD sum rules.
Unfortunately, the sum rules obtained in refs. [10-13]
differ in nonleading terms and, to some extent, also in numerical results.
Here, we suggest
an alternative method known as QCD sum rules on
the light-cone. In this approach, the ideas of duality and matching
between parton and hadron descriptions, intrinsic to the
QCD sum rules, are combined with the specific operator
product expansion (OPE) techniques used to study hard exclusive
processes in QCD \cite{exclusive,BLreport}.
In contrast to the conventional sum rules based on
the Wilson OPE of the T-product of currents at small
distances, one considers expansions near the light-cone
in terms of nonlocal operators, the matrix elements of which define hadron
wave functions of increasing twist. As one advantage, this
formulation allows to incorporate additional
information about the Euclidean asymptotics of correlation functions
in QCD  for arbitrary external momenta. These new features are
related to the (approximate) conformal invariance of QCD
and are coded in the hadron wave functions. Many of the theoretical
results obtained in the context of exclusive processes
(see e.g. ref. \cite{CZreport})  are very useful
in the present context as well.
In turn, we will see that heavy-flavour decays can provide
valuable constraints on  the wave functions.

Previous applications of light-cone sum rules include
calculations of the amplitude of the radiative decay
$\Sigma\rightarrow p\gamma$ \cite{BBK}, the nucleon magnetic moments
\cite{BF1}, the strong couplings $g_{\pi N N}$ and $g_{\rho\omega\pi}$
\cite{BF1}, form factors of semileptonic and radiative
$B$- and $D$-meson decays [8,19-21],
the pion form factor at
intermediate momentum transfers \cite{BH},  and the $ \pi A \gamma^* $
form factors \cite{belyaev94}. In all these cases the results
are encouraging.

The light-cone sum rule for the coupling of heavy mesons to a pion
is the principal result of the present paper which is organized as follows.
In Sect.~2 we discuss possible strategies in constructing
sum rules for coupling constants and explain the concept of
light-cone sum rules. The derivation of the sum rule
for the $B^*B\pi$ and $D^*D\pi$ couplings is then completed in Sect.~3,
taking into account the pion two- and three-particle wave
functions up to twist 4. Sect.~4 is devoted to a detailed
numerical analysis. In Sect.~5 we show that in a
simplified case, putting the external momenta in the correlation function
equal to each other and performing a Borel transformation in one
momentum instead of
two, we obtain the sum rule proposed previously in refs. [10-13].
We demonstrate that despite the slightly different terminology
of these papers the sum rules must coincide with each
other, and elaborate on possible subtleties in these earlier calculations.
Furthermore, in Sect. 6 using our results on the
$B^*B\pi$ and $D^*D\pi$ coupling
constants, we confront the pole model for the heavy-to-light form factors
$B \rightarrow \pi$ and $D \rightarrow \pi$ with a direct calculation
of these form factors  in the same framework of
light-cone sum rules following ref. \cite{BKR}.
A comprehensive comparison of our results on the $ D^*D\pi$ and $B^*B\pi$
couplings with other estimates and our conclusions are presented in Sect.~7.

Technical details are collected
in two Appendices.
Appendix A summarizes the
relevant features of the pion wave functions and specifies the input
in our numerical calculations.
In Appendix B we derive a simple rule how to subtract
the contribution from
excited resonances and continuum states in the sum rule.


\section{ Light-cone versus short-distance expansion}

For definiteness, we focus on the
$D^{*}D \pi$ coupling defined by the on-mass-shell matrix element
\bq
\langle D^{*+}(p)\pi^-(q)\mid D^0(p+q)\rangle &=&
-g_{D^*D\pi}q_\mu \epsilon ^{\mu},
\label{1}
\eq
where the momentum assignment is specified in brackets
and
 $\epsilon_\mu$ is the polarization vector of the $D^*$.
The couplings for the different charge states are related by isospin
symmetry:
\bq
g_{D^*D\pi} \equiv g_{D^{*+}D^0\pi^+}=
-\sqrt{2}g_{D^{*+}D^{+}\pi^0}=\sqrt{2}g_{D^{*0}D^0\pi^0}
=-g_{D^{*0}D^+\pi^-}~.
\label{gd}
\eq
Most of what is said below applies equally to $B^*B\pi$ couplings.
The corresponding relations are
obtained by the obvious replacements $ c \ra b $,
$ D^* \ra \bar B^*$ and $D \ra \bar B $.

Following the general strategy of QCD sum rules, we want to
obtain quantitative estimates for $g_{D^*D\pi}$ by matching
the representations of a suitable correlation function in terms
of hadronic and quark-gluon degrees of freedom. For this purpose,
we choose
\be
F_\mu (p,q)=
i\int d^4xe^{ipx} \langle\pi^-(q)|T\{ \bar{d}(x)\g_\mu c(x),
\bar{c}(0)i\g_5 u(0)\} |0\rangle  ~.
\label{19}
\ee
To the best of
our knowledge, the study of
correlation functions with the T-product of currents sandwiched
between the vacuum and one-pion state was first suggested in
ref. \cite{CS83}.

With the pion being on mass-shell, $q^2=m_\pi^2 $, the
correlation function (\ref{19})
depends on two invariants, $p^2$ and $(p+q)^2$.
Throughout the paper we set $m_\pi=0$.
The contribution of interest
is the one having poles in $p^2$ and $(p+q)^2$ :
\bq
F_\mu (p,q)& = &\frac{m_D^2m_{D^*}f_Df_{D^*}g_{D^*D\pi}}{m_c
(p^2-m_{D^*}^2)((p+q)^2-m_D^2)}
(q_\mu+\frac12(1-\frac{m_D^2}{m_{D^*}^2})p_\mu)~.
\label{22}
\eq
Obviously, this term stems from the ground states in the
$(\bar{d}c)$ and $(\bar c u)$ channels. To derive eq. (\ref{22})
we have made use of eq. (\ref{1}) and the decay constants $f_D$
and $f_{D^*}$ defined by the matrix elements
\bq
\langle D \mid \bar{c}i\g_5u \mid 0\rangle &=&\frac{m_D^2f_D}{m_c}
\label{fD1}
\eq
and
\bq
\langle 0 \mid \bar{d}\g_\mu c\mid D^*\rangle
&=&m_{D^*}f_{D^*}\epsilon_\mu   ~,
\label{fD*}
\eq
respectively.

The main theoretical task is the calculation of the correlation
function (\ref{19}) in QCD. This problem can be solved
in the Euclidean region
where both virtualities  $p^2$ and $(p+q)^2$ are negative
and large, so that the charm quark is sufficiently far off-shell.
Substituting, as a first approximation,
the free $c$-quark propagator
\bq
\langle 0|T\{c(x)\bar{c}(0)\}|0\rangle = i\hat{S}_c^0(x) =
\int \frac{d^4k}{(2\pi)^4i}e^{-ikx}
\frac{\not\!k+m_c}{m_c^2-k^2}
\label{prop}
\eq
into eq. (\ref{19}) one readily obtains
\bq
F_\mu(p,q)&=&i\int \frac{
d^4x\,d^4k}{(2\pi )^4(m_c^2-k^2)}
e^{i(p-k)x}\left(m_c
\langle \pi (q)|\bar{d}(x)\g_\mu\g_5u(0)|0\rangle \right.
\nonumber
\\
&&{}+\left.
k^\nu \langle\pi(q) |\bar{d}(x)\g_\mu\g_\nu\g_5u(0)|0\rangle\right)~.
\label{23}
\eq
Diagramatically, this contribution is depicted in Fig.~1a.
Applying the short-distance expansion (SDE)
in terms of local operators to the first matrix element
of eq. (\ref{23}),
\be
\bar{d}(x)\g_\mu\g_5u(0)=\sum_n\frac{1}{n!}
\bar{d}(0)(\stackrel{\leftarrow}{D}\cdot x)^n\g_\mu\g_5 u(0)~,
\label{expan}
\ee
one has after integration over $x$ and $k$ :
\bq
F_\mu(p,q)&=&i\frac{m_c}{m_c^2-p^2}
\sum_{n=0}^\infty \frac{(2p \cdot q)^n}{(m_c^2-p^2)^n}M_n q_\mu ~,
\label{expans}
\eq
where
$$\langle\pi(q) |\bar{d}\stackrel{\leftarrow}{D}_{\alpha_1}
\stackrel{\leftarrow}{D}_{\alpha_2}...
\stackrel{\leftarrow}{D}_{\alpha_n}\g_\mu\g_5 u  |0\rangle
=(i)^n q_\mu q_{\alpha_1} q_{\alpha_2}...q_{\alpha_n}M_n + ...~, $$
$D$ being the covariant derivative, has been used.
One immediately encounters the following problem. If
the ratio
\bq
\tilde{\xi}=2(p \cdot q)/(m_c^2-p^2)= ((p+q)^2-p^2)/(m_c^2-p^2)
\label{ksi}
\eq
is finite one must keep an {\em infinite} series of local operators
in eq. (\ref{expans}). All these operators give
contributions of the same order in the heavy quark propagator
$1/(m_c^2-p^2)$, differing only by powers
of the dimensionless parameter $ \tilde{\xi}$
\footnote {
This feature is also observed in deep
inelastic scattering, with the variables $\{ Q^2, ~\nu, ~x \}$
playing the role of
$ \{-p^2,~ p\cdot q ,~ \tilde{\xi} \}$ . As well known, there one applies
an expansion near the light-cone in terms of operators of increasing twist,
rather than of increasing dimension. }. Therefore,
SDE of eq. (\ref{23}) is useful {\em only} if $ \tilde{\xi} \rightarrow 0$,
i.e. if $p^2 \simeq (p+q)^2$
or, equivalently, $q \simeq 0$. Under this condition, the series in
eq. (\ref{expans}) can be truncated after a few terms involving only a small
number of unknown matrix elements $M_n$. However, for general momenta with
$p^2 \neq (p+q)^2$ one has to sum up the infinite series of matrix elements
of local operators in some way.

This formidable task is solved by using
the techniques developed for hard exclusive processes in QCD
\cite{BLreport,CZreport}. We illustrate the solution for the
correlation function
\bq
i\int d^4x\,e^{ipx}\langle\pi^0(q)|T\{ \bar{q}(x)\g_\mu {\cal Q}q(x),
\bar{q}(0)\g_\nu{\cal Q} q(0)\} |0\rangle =
\varepsilon_{\mu\nu\alpha\beta} p^\alpha q^\beta F^*(p^2,(p+q)^2)~,
\label{photonff}
\eq
which is similar to eq. (\ref{19}) and
defines the form factor of the coupling of a pion to a pair of
virtual photons \cite{CZreport,gorsky}.
In eq. (\ref{photonff}), ${\cal Q}$ is a matrix of electromagnetic charges,
and $q$ is a row vector composed of the up and down quark flavours.
As well-known \cite{exclusive}, for sufficiently
virtual photons, $p^2 \ra -\infty$ and $(p+q)^2 \ra -\infty$, this
form factor can be calculated in perturbative QCD. The principal result reads
\bq
F^*(p^2,(p+q)^2) = F_0 \int_0^1 \frac{du \,\vp_\pi(u)}
  {(p+uq)^2}\,,~~~~~F_0=\frac{4\pi\alpha\sqrt{2}f_\pi}{3} ~,
\label{photon}
\eq
with calculable radiative corrections, and with power corrections
suppressed by the photon virtualities.
Here, $\vp_\pi(u)$ is the pion wave function of leading
twist, defined by the following matrix element of a nonlocal operator
on the light-cone $x^2=0$ :
\bq
\langle\pi(q)|\bar{d}(x)\g_\mu\g_5u(0)|0\rangle=
-iq_\mu f_\pi\int_0^1du\,e^{iuqx}\vp_\pi (u)  ~.
\label{pionwf}
\eq
Physically, $\vp_\pi$ represents the distribution in the fraction
of the light-cone momentum $q_0 +q_3 $ of the pion carried by
a constituent quark. Note the normalization of $\vp_\pi$ to unity
following from eq. (\ref{pionwf}) for $x=0$ .

Let us first concentrate on the
form factor (\ref{photon}) at (almost) equal
photon virtualities, i.e. at
$\xi= (2p\cdot q)/(-p^2) \ll 1$.
Expanding the denominator in eq. (\ref{photon})
around $\xi=0$ one obtains a sum over
moments of the pion wave function:
\be
F^*(p^2,(p+q)^2)= \frac{F_0}{p^2}\sum _n
\xi^n\int_0^1du\,u^n\vp_\pi (u) ~.
\label{summom}
\ee
{}~From the definition (\ref{pionwf}) it is easy to
see that these moments are given by vacuum-to-pion transition
matrix elements involving increasing powers of the covariant
derivative. For $ p^2=(p+q)^2 $, i.e. $q=0$, only the lowest moment $n=0$
contributes in eq. (\ref{summom}), and the form factor
reduces to $F_0/p^2$ which is the
classical result. In contrast, if the photon virtualities differ
strongly from each other, then many moments contribute to eq. (\ref{summom}).
In this case, the calculation of the form factor requires the
knowledge of the shape of the pion wave function.

Returning to the correlation function (\ref{19}) one realizes that
the same technique may be used to obtain a representation
analogous to eq. (\ref{photon}).
The only new element in the correlation function
(\ref{19}) is the virtual heavy quark propagating between the points
$x$ and $0$ instead of the light quarks present in eq. (\ref{photonff}).
This gives rise to important differences which however do not change
the formalism substantially. For the present discussion it
is sufficient to stick to the approximation (\ref{23}) and confine
ourselves to the first term proportional to $m_c$.
The complete analysis of this expression and the calculation of
further corrections will be carried out in the next section.
Furthermore, writing $F_\mu$  in terms of
invariant amplitudes:
\bq
F_\mu (p,q)=F(p^2,(p+q)^2)q_\mu + \tilde{F}(p^2,(p+q)^2)p_\mu ~,
\label{220}
\eq
we focus on the function $F$.
Using the definition eq. (\ref{pionwf}) of the
leading twist wave  function
and integrating over $x$ and $k$ one finds
\be
F(p^2,(p+q)^2)=m_cf_\pi\int_0^1\frac{du ~\vp_\pi(u) }{m_c^2-(p+uq)^2} ~.
\label{Fzeroth}
\ee
Thus, the infinite series of matrix elements of local operators
encountered before
in eq. (\ref{expans})  is effectively replaced by an unknown wave function.
The expression (\ref{Fzeroth})
is rather similar to the one quoted in eq. (\ref{photon})
for the $ \pi^0 \g ^* \g^* $ form factor. Most noteworthy is the fact that
the large-distance dynamics is described by one and the same
pion wave function. This universal property is essential for the whole
approach.

Next we indicate how the relation (\ref{Fzeroth}) can be turned into a
sum rule for the coupling constant $g_{D^*D\pi}$.
The key idea is to write a hadronic representation of $F$ by means
of a double dispersion integral:
\bq
F(p^2,(p+q)^2)=\frac{m_D^2m_{D^*}f_Df_{D^*}g_{D^*D\pi}}{m_c
(p^2-m_{D^*}^2)((p+q)^2-m_D^2)}
\nonumber
\\
+\int\frac{\rho^h(s_1,s_2)ds_1ds_2}{(s_1-p^2)(s_2-(p+q)^2)}
\nonumber
\\
+\int\frac{\rho ^h _1(s_1)ds_1}{s_1-p^2}
+\int\frac{\rho ^h _2(s_2)ds_2}{s_2-(p+q)^2}~.
\label{221}
\eq
The first term arises from the ground state contribution already indicated
in eq. (\ref{22}), while  the spectral function $\rho^h(s_1,s_2)$
is supposed to take into
account higher resonances and continuum states in the
$D^*$ and $D$ channels. The additional single dispersion integrals
originate in subtractions which are generally necessary to make the
double dispersion integral finite. Then, considering
$p^2$ and $(p+q)^2$ as independent
variables one can perform the usual Borel improvement in both channels.
Applying the Borel operator
\bq
{\cal B}_{M^2}f(Q^2)=lim_{Q^2,n \ra \infty , Q^2/n=M^2 }
\frac{(Q^2)^{(n+1)}}{n!}\left( -\frac{d}{dQ^2}\right)^n f(Q^2)
\equiv f(M^2)
\label{B}
\eq
to eq. (\ref{221}) with respect to
$p^2$ and $(p+q)^2$, we obtain
\bq
F(M_1^2,M_2^2)&\equiv&
{\cal B}_{M_1^2}{\cal B}_{M_2^2}F(p^2,(p+q)^2)=
\frac{m_D^2m_{D^*}f_Df_{D^*}g_{D^*D\pi} }{m_c}
e^{-\frac{m_{D^*}^2}{M_1^2}-\frac{m_{D}^2}{M_2^2}}
\nonumber
\\
&+&\int e^{-\frac{s_1}{M_1^2}-\frac{s_2}{M_2^2}}
\rho^h(s_1,s_2)ds_1ds_2  ~,
\label{222}
\eq
where  $M_1^2$ and $M_2^2$  are the Borel parameters associated with
$p^2$ and $(p+q)^2$, respectively.
Note that
contributions from heavier states are now exponentially suppressed
by factors
$\exp\{-\frac{s_{1,2}^2-m_{D^*,D}^2}{M_{1,2}^2} \}$ as desired, while
 the subtraction terms depending only on one of the
variables, $p^2$ or $(p+q)^2$, vanish.

The same transformation has to be applied to
the expression (\ref{Fzeroth}). To this end we rewrite
$(p+uq)^2 = (1-u) p^2 + u (p+q)^2$, and use
\begin{equation}
{\cal B}_{M_1^2}{\cal B}_{M_2^2}
\frac{(l-1)!}{[m_c^2 -(1-u) p^2 - u (p+q)^2]^l }
 = (M^2)^{2-l} e^{-m_c^2/M^2 }\delta(u-u_0) ~,
\label{dBorel}
\end{equation}
where the Borel parameters $M_1^2$ and $M^2_2$ have been replaced
by
\be
u_0=\frac{M_1^2}{M^2_1+M_2^2}, ~~
M^2=\frac{M_1^2M^2_2}{M^2_1+M_2^2}  ~.
\label{37}
\ee
Finally, equating the quark-gluon and the hadronic representations of
$F(M_1^2,M_2^2) $
and discarding for a moment contributions of higher states,
we end up with the  sum rule
\begin{equation}
\frac{m_D^2m_{D^*}f_Df_{D^*}}{m_c} \cdot g_{D^*D\pi}
= m_c f_\pi~\vp_\pi(u_0)M^2
\exp \left[ \frac{m^2_{D^\ast}-m_c^2}{M_1^2} +
\frac{m^2_{D}-m_c^2}{M_2^2}\right] +\ldots
\label{SR0}
\end{equation}
The ellipses refer to higher-twist
contributions which we discuss in detail later.
Since $M_1^2$ and $M_2^2$ are expected to be quite similar in magnitude,
the coupling constant $g_{D^*D\pi}$ is determined by the value of the
pion wave function at $u\simeq 1/2$,
that is by the probability for the quark and
the antiquark to carry equal momentum fractions in the pion \cite{BBK}.
This interesting feature is shared by the sum rules for many other
important hadronic couplings involving the pion.

The quantity $\vp_\pi(1/2)$  is considered to be
a nonperturbative parameter, similar to
quark and gluon condensates in the standard approach. It
may be determined from suitable sum rules in which the
phenomenological part is known experimentally. We use the value
\begin{equation}
    \vp_\pi(1/2) = 1.2\pm 0.2
\label{WF12}
\end{equation}
obtained in ref. \cite{BF1}.

The dependence on the pion wave function disappears in
the kinematical limit $q\rightarrow 0$ as can be seen from
eq. (\ref{Fzeroth}). This is just the limit where the correlation
function (\ref{19}) can be treated in SDE.
The condition $q \simeq 0$
is implicitly assumed in refs.
\cite{EK85,GY94} where the correlation function (\ref{19})
is calculated using the external field method. This technique is
equivalent to the soft-pion approximation used in refs. \cite{ovch,Cetc94}
as will become clear later.
For comparison, we present the sum rule following from
eqs. (\ref{Fzeroth}) and (\ref{221})
by putting $q=0$, or equivalently $(p+q)^2=p^2$.
Since $p^2$ is the only variable left, one now can only perform a single
Borel transformation and, hence, the subtraction terms in the double
dispersion relation (\ref{221})
are no more eliminated.
Moreover, the contributions to eq. (\ref{221})
of transitions from excited states to ground states are not suppressed after
Borel transformation \cite{ioffe}.
This point will be explained in more detail later in Sect. 5 .
In the approximation considered  in eq. (\ref{SR0}) one obtains
\begin{equation}
\frac{m_D^2m_{D^*}f_Df_{D^*}}{m_c} \cdot g_{D^*D\pi} + M^2 A
= m_c f_\pi M^2
\exp \left[ \frac{m^2_{D^\ast}-m_c^2}{2M^2} +
\frac{m^2_{D}-m_c^2}{2M^2}\right] +\ldots  ~,
\label{SR1}
\end{equation}
where $A$ is an unknown constant corresponding to the
contributions of unwanted transitions and subtraction terms.

{}~From eqs. (\ref{SR0}) and (\ref{SR1}) one can clearly see the
advantages and disadvantages of the two approaches. In the
light-cone sum rule (\ref{SR0}) the hadronic input is simple, whereas
the theoretical expression involves a new
{\em universal} nonperturbative parameter, namely  $\vp_\pi(1/2)$.
Just the opposite is the case for the sum rule (\ref{SR1}) at $q =0$.
Here the QCD part is straightforward, while the hadronic representation
now involves an additional
unknown quantity, which is {\em non-universal} and specific for this
particular sum rule.
A comparison of the results obtained in these two approaches
should allow one to check
the reliability and improve the accuracy of the predictions.

\section{ Light-cone sum rule for $g_{D^*D\pi}$ and $g_{B^*B\pi}$}

In this section we systematically  derive the light-cone sum rule for
the $D^*D\pi$ and $B^*B\pi$ couplings taking into account the two-
and three-particle pion wave functions up to twist 4.
First, we complete the calculation of the diagram
Fig.~1a which represents the contribution from quark-antiquark wave
functions. To this end we return to the expression (\ref{23}). In
the first matrix element we include
the twist 4 corrections in addition to the leading
twist term already given in eq. (\ref{pionwf}):
\begin{eqnarray}
\langle\pi(q)|\bar{d}(x)\g_\mu\g_5u(0)|0\rangle &=&
-iq_\mu f_\pi\int_0^1du\,e^{iuqx}
\left(\vp_\pi (u)+x^2g_1(u)+O(x^4)\right)
\nonumber \\
&&{}+
f_\pi\left( x_\mu -\frac{x^2q_\mu}{qx}\right)\int_0^1
du\,e^{iuqx}g_2(u) ~.
\label{25}
\end{eqnarray}
On the r.h.s. of this relation one sees the first few terms of the light-cone
expansion in $x^2$ of the matrix element on the l.h.s.. While
$\vp_\pi$ parametrizes the leading twist 2 contribution, $g_1$
and $g_2$ are associated with twist 4 operators.
In the second matrix element of eq. (\ref{23}) we substitute
\be
\g_\mu\g_\nu=-i\sigma_{\mu\nu}+g_{\mu\nu}
\label{gammaid}
\ee
and express the result
in terms of the twist 3 wave functions $\vp_p$ and
$\vp_\sigma $ defined by the matrix elements
\be
\langle\pi (q)|\bar{d}(x)i\g_5u(0)|0\rangle
=\frac{f_\pi m_\pi^2}{m_u+m_d}
\int_0^1du\,e^{iuqx}\vp_{p}(u)
\label{26}
\ee
and
\be
\langle\pi (q)|\bar{d}(x)\sigma_{\mu\nu}\g_5u(0)|0\rangle=
i(q_\mu x_\nu -q_\nu x_\mu )\frac{f_\pi m_\pi^2}{6(m_u+m_d)}
\int_0^1du\,e^{iuqx}\vp_\sigma (u)
\label{27} ~.
\ee
It should be noted that in eqs. (\ref{25},\ref{26},\ref{27})
the path-ordered gauge factors
\be
\mbox{\rm Pexp} \{ ig_s \int^1_0 d \alpha x_\mu A^\mu (\alpha x) \} ~,
\label{path}
\ee
appearing
in between the quark fields and assuring gauge invariance,
are not shown for brevity since they formally disappear in
the light-cone gauge $x_\mu A^\mu =0 $ assumed throughout this paper.
More details on these wave functions can be found in
refs. \cite{BF1,BB,BF2} and in Appendix A.

Collecting all terms , we obtain the following result for the
invariant function $F$ as defined in eq. (\ref{220}):
\bq
F^{(a)}(p^2,(p+q)^2)&=&\int_0^1\frac{du}{m_c^2-(p+uq)^2}\left\{
m_cf_\pi \vp_\pi (u)+\frac{f_\pi m_\pi^2}
{m_u+m_d}\Bigg[ u\vp_p(u)\right.
\nonumber
\\
&&{}
+\left.\left. \frac16 \left(
2+\frac{p^2+m_c^2}{m_c^2-(p+uq)^2}\right)\vp_\sigma (u)
\right]\right.
\nonumber
\\
&&{}
+\left. m_cf_\pi \left[\frac{2ug_2(u)}{m_c^2-(p+uq)^2}-
\frac{8m_c^2(g_1(u)+G_2(u))}{(m_c^2-(p+uq)^2)^2}\right]\right\}~,
\label{28}
\eq
where $$G_2(u)=-\int_0^ug_2(v)dv ~.$$
The suffix $(a)$ refers to the diagram in Fig. 1a which represents
the leading twist term
in the light-cone expansion of the c-quark propagator given in
eq. (\ref{prop}).

In addition, to the accuracy of eq. (\ref{28})
we must also take into account higher
twist terms in the propagator up to twist 4
which are numerous, in general. The complete expansion is given in
ref. \cite{BB}. One has contributions from
$\bar q G q$, $\bar q GG q $ and $\bar q q\bar q q$ nonlocal operators,
$G$ denoting the gluon field strength. Here we only consider
operators with one gluon field, corresponding to quark-antiquark-gluon
components in the pion, and neglect components with two extra gluons,
or with an additional $\bar q q$ pair. This is consistent with the
approximation of the
twist 4 two-particle wave functions derived in ref. \cite{BF2}
and used here.
Taking into account higher Fock-space  components
would demand corresponding modifications
in the two-particle functions via the equations of motion.
Formally, the neglect of the $\bar q GG q $ and $\bar q q\bar q q$
terms can be justified on the basis of an
expansion in conformal spin \cite{BF2}.
In this approximation the $c-$quark propagator reads
\begin{eqnarray}
\langle 0 |T\{c(x)\bar{c}(0)\}|0\rangle &=& i\hat{S}_c^0(x)
-ig_s\int\frac{d^4k}{(2\pi )^4}e^{-ikx}
\int_0^1dv\left[ \frac12\frac{\not\!k+m_c}{(m^2_c-k^2)^2}
G^{\mu\nu}(vx)\sigma_{\mu\nu}\right.
\nonumber\\
&& \mbox{}
+\left.\frac1{m_c^2-k^2}vx_\mu G^{\mu\nu}(vx)\g_\nu
\right]~,
\label{32}
\end{eqnarray}
where $G_{\mu\nu}=G_{\mu\nu}^c\frac{\lambda^c}2$ with
$\mbox{\rm tr}(\lambda^a\lambda^b)=2\delta^{ab}$, and $g_s$ is the
strong coupling constant.

Substituting eq. (\ref{32}) into eq. (\ref{19}) and using
eq. (\ref{220})
one obtains
the contribution to the invariant function $F$
represented by the diagram in Fig. 1b:
\bq
F_\mu^{(b)}
(p^2, (p+q)^2)&=&i\int\frac{d^4k\,d^4x\,dv}{(2\pi)^4(m_c^2-k^2)}
\langle \pi |\bar{d}(x)\g_\mu
\Bigg[ vx_\rho G^{\rho\lambda}(vx)\g_\lambda
\nonumber
\\
&&{}+\frac{\hat{k}+m_c}{m_c^2-k^2}\frac12G^{\rho\lambda}(vx)\sigma_{\rho
\lambda}\Bigg]\g_5u(0)|0\rangle~.
\label{33}
\eq
 With  eq. (\ref{gammaid}) and the identities
\bq
\g_\mu\sigma_{\rho\lambda}&=&
i(g_{\mu\rho}\g_\lambda-g_{\mu\lambda}\g_\rho)
+\varepsilon_{\mu\rho\lambda\delta}\g^\delta\g_5
\label{ggg}
\eq
and
\bq
\g_\mu\g_\nu\sigma_{\rho\lambda}&=&
(\sigma_{\mu\lambda}g_{\nu\rho}-\sigma_{\mu\rho}g_{\nu\lambda})+
i(g_{\mu\lambda}g_{\nu\rho}-g_{\mu\rho}g_{\nu\lambda})
\nonumber
\\
&&\mbox{}-
\varepsilon_{\mu\nu\rho\lambda}\g_5-i\varepsilon_{\nu\rho\lambda\al}
g^{\al\beta}
\sigma_{\mu\beta}\g_5
\label{34}
\eq
one is led to the three-particle pion wave
functions \cite{gorsky,BF2}  defined by
\bq
\lefteqn{
\langle\pi |\bar{d}(x)
g_s G_{\mu\nu}(vx)\sigma_{\al\bet}\g_5u(0)|0\rangle =}
\nonumber\\
&=&if_{3\pi}[(q_\mu q_\al g_{\nu\bet}-q_\nu q_\al g_{\mu\bet})
-(q_\mu q_\bet g_{\nu\al}-q_\nu q_\bet g_{\mu\al})]
\int{\cal D}\al_i\,\vp_{3\pi}(\al_i)e^{iqx(\al_1+v\al_3)}\,,
\label{29}
\eq
\bq
\lefteqn{
\langle\pi |\bar{d}(x)\g_\mu\g_5 g_sG_{\al\bet}(vx)u(0)|0\rangle =}
\nonumber
\\
&=&f_\pi\left[ q_\bet\left( g_{\al\mu}-\frac{x_\al q_\mu}{qx}\right) -
q_\al\left(g_{\bet\mu}-\frac{x_\bet q_\mu}{qx}\right)\right]
\int{\cal D}\al_i\vp_\perp (\al_i)e^{iqx(\al_1+v\al_3)}\hspace{1.5cm}{}
\nonumber
\\
&& {}+f_\pi\frac{q_\mu}{qx}(q_\al x_\bet -q_\bet x_\al )
\int{\cal D}\al_i\,\vp_\parallel (\al_i)e^{iqx(\al_1+v\al_3)}\,,
\label{30}
\eq
\bq
\lefteqn{
\langle\pi |\bar{d}(x)\g_\mu g_s\tilde{G}_{\al\bet}(vx)u(0)|0\rangle=}
\nonumber
\\
&=&if_\pi\left[ q_\bet\left( g_{\al\mu}-\frac{x_\al q_\mu}{qx}\right) -
q_\al\left( g_{\bet\mu}-\frac{x_\bet q_\mu}{qx}\right)\right]
\int{\cal D}\al_i\,\tilde{\vp}_\perp (\al_i)e^{iqx(\al_1+v\al_3)}
\hspace{1.4cm}{}
\nonumber
\\
&&{}+if_\pi\frac{q_\mu}{qx}(q_\al x_\bet -q_\bet x_\al )
\int{\cal D}\al_i\,\tilde{\vp}_\parallel (\al_i)e^{iqx(\al_1+v\al_3)}\,,
\label{31}
\eq
where $\tilde{G}_{\al\bet}= \frac12
\epsilon _{\al\bet \sigma\tau}G^{\sigma\tau}$  and
${\cal D}\al_i= d\al_1 d\al_2 d\al_3 \delta(1-\al_1-\al_2-\al_3)$.
While $\vp_{3\pi}(\al_i)$ is a twist 3 wave function, the remaining
functions
$\vp_\perp$, $\vp_\parallel$,
$\tilde{\vp}_\perp$ and $\tilde{\vp}_\parallel$ are all of twist 4.
Substitution of these expressions
in eq. (\ref{33}), finally, yields
\bq
F^{(b)}(p^2,(p+q)^2)&=& \int_0^1 dv\int{\cal D}\al_i\left\{
\frac{4f_{3\pi}\vp_{3\pi}(\al_i)v(pq)}
{[m_c^2-(p+(\al_1+v\al_3)q)^2]^2}\right.
\nonumber
\\
&&{}+ m_cf_\pi
\left.\frac{2\vp_\perp (\al_i)-\vp_\parallel (\al_i)+
2\tilde{\vp}_\perp (\al_i)-\tilde{\vp}_\parallel (\al_i)}
{[m_c^2-(p+(\al_1+v\al_3)q)^2]^2}\right\} ~.
\label{35}
\eq

In addition to Fig. 1b there are further gluonic diagrams
such as the ones depicted in Figs. 1c and 1d.
Note, however, that it is not necessary to take
the diagram in Fig.~1c into account separately,
since its contribution (to twist 4 accuracy)
is already included in the two particle wave functions.
 In contrast, the two-loop perturbative
corrections exemplified
in Fig.~1d should be included in a
systematic way, but their calculation lies beyond the scope of this
paper.

Putting together eqs. (\ref{28}) and (\ref{35}) and applying the
double Borel transformation (\ref{dBorel})
with respect to $p^2$ and $(p+q)^2$, we end up with the following
expression for the invariant amplitude $F$ :
\bq
\lefteqn{
F(M_1^2,M_2^2)\equiv F(M^2,u_0)\equiv
{\cal B}_{M_1^2}{\cal B}_{M_2^2}F(p^2,(p+q)^2)= }
\nonumber\\
&=&
e^{-\frac{m_c^2}{M^2}}M^2 \Bigg\{ m_cf_\pi\vp_\pi(u_0)
+\frac{f_\pi m_\pi^2}{m_u+m_d} \left( u_0 \vp_p(u_0)
+\frac13\vp_\sigma (u_0)+\frac16u_0 \frac{d\vp_\sigma}{du}(u_0)
+\frac{m_c^2}{3 M^2}\vp_\sigma (u_0) \right)
\nonumber
\\&&{}
+\frac{2f_\pi m_c}{M^2}u_0g_2(u_0)-\frac{4f_\pi m_c^3}{M^4}
(g_1(u_0) +G_2(u_0)) + 2f_{3\pi}I^G_3(u_0)+
m_cf_\pi\frac{I^G_4(u_0)}{M^2} \Bigg\}  ~.
\label{F}
\eq
Here, $I^G_{3}$ and $I^G_{4}$ involve the
three-particle wave functions of twist 3 and 4, respectively :
\be
I^G_3(u_0)=
\int_0^{u_0}d\al_1 \left[\frac{\vp_{3\pi}(\al_1,1-u_0,u_0-\al_1)}
{u_0-\al_1}
-\int_{u_0-\al_1}^{1-\al_1}d\al_3
\frac{\vp_{3\pi}(\al_1,1-\al_1-\al_3,\al_3)}{\al_3^2}\right]~,
\label{fg3}
\ee
\be
I^G_4(u_0)=
\int_0^{u_0}d\al_1\int_{u_0-\al_1}^{1-\al_1}
\frac{d\al_3}{\al_3}[2\vp_\perp (\al_i)-\vp_\parallel (\al_i)+
2\tilde{\vp}_\perp (\al_i)-\tilde{\vp}_\parallel (\al_i)] ~.
\label{fg4}
\ee
The Borel parameters $M^2$ and $u_0 $ are
given  in eq. (\ref{37}). The above
is the desired quark-gluon representation of the
invariant amplitude $F$ in the correlation function (\ref{19}).

The remaining task now is to match eq. (\ref{F}) with the corresponding
hadronic representation (\ref{222}) and to extract the
coupling $g_{D^*D\pi}$. As usual, invoking duality, we assume that
above certain thresholds in $s_1$ and $s_2$ the
double spectral density $\rho^h(s_1,s_2)$ associated with higher resonances
and continuum states coincides
with the spectral density derived from the diagrams in Fig.~1.
The procedure is explained in detail in Appendix~B.
For $M_1^2=M_2^2=2M^2$ and
$u_0=\frac12$, and for standard polynomial
wave functions, the effect of the continuum
subtraction is remarkably simple \cite{BBK,BF1}.
It amounts to the following
replacement of the exponential factor multiplying the twist 2
and 3 terms proportional to $M^2$ in eq. (\ref{F}):
\bq
e^{-\frac{m_c^2}{M^2}}\ra \left( e^{-\frac{m_c^2}{M^2}}-
e^{-\frac{s_0}{M^2}}\right)~,
\label{39}
\eq
$s_0$ being the threshold parameter defined in eq. (\ref{boundary}).
The higher twist terms which are suppressed in eq. (\ref{F})
 by inverse powers of $M^2$
with respect to leading one remain  unaffected.
With eqs. (\ref{222}), (\ref{F}) and (\ref{39})
it is then easy to derive the
following QCD sum rule for the $DD^*\pi$ coupling:
$$
f_Df_{D^*}g_{D^*D\pi}=\frac{m_c^2f_\pi}{m_D^2m_D^*}
e^{\frac{m_{D}^2+m_{D*}^2}{2M^2}}
\Bigg\{M^2 [e^{-\frac{m_c^2}{M^2}} - e^{-\frac{s_0}{M^2}}]
\Big[\vp_\pi(u_0)
$$
$$
+ \frac{\mu_\pi}{m_c} \left( u_0\vp_p(u_0)
+\frac13\vp_\sigma (u_0)+\frac16u_0\frac{d\vp_\sigma}{du}(u_0)\right)
+ \frac{2f_{3\pi}}{m_cf_\pi}I^G_3(u_0)\Big]
$$
\be
+e^{-\frac{m_c^2}{M^2}}
\Big[\frac{\mu_\pi m_c}3 \vp_\sigma (u_0)
+2u_0g_2(u_0)-\frac{4m_c^2}{M^2}
(g_1(u_0)+G_2(u_0)) + I^G_4(u_0)\Big]\Bigg\}_{u_0=1/2}~,
\label{fin}
\ee
where
\begin{equation}
 \mu_\pi=\frac{m_\pi^2}{m_u+m_d}
 = \frac{-2\langle\bar q q\rangle}{f_\pi^2} ~.
\label{qqbar}
\end{equation}
In eq. (\ref{qqbar}) we have used the familiar PCAC relation between
$m_\pi$, $f_\pi$ , the quark masses and the quark condensate
density $\langle\bar q q\rangle$.
Note that the twist 2 and 3 and the twist 4 wave functions
have different dimensions (see Appendix~A). $G$-parity implies
$g_2(1/2)=d\vp_\sigma/du(1/2)=0$, so that these terms vanish in the
sum rule (\ref{fin}).

For completeness, we also repeat the standard two-point sum rules
for the decay constants $f_D$ and $f_{D^\ast}$ :
\be
\frac{f_D^2m_D^4}{m_c^2}=
\frac{3}{8\pi^2}
\int_{m_c^2} ^{s_0}ds e^{\frac{m_D^2-s}{M^2}}\frac{(s-m_c^2)^2}s
-m_c \langle \bar q q \rangle e^{\frac{m_D^2-m_c^2}{M^2}}\left[1+
\frac{m_0^2}{2M^2}(1-\frac{m_c^2}{2M^2})\right]
 ~,
\label{fD}
\ee
and
\be
f_{D^*}^2m_{D*}^2=
\frac{1}{8 \pi^2}\int_{m_c^2} ^{s_0} ds
e^{\frac{m_{D^*}^2-s}{M^2}}\frac{(s-m_c^2)^2}s
\left(2+\frac{m_c^2}{s}\right)
-m_c \langle \bar q q \rangle e^{\frac{m_{D^*}^2-m_c^2}{M^2}}
\left(1-\frac{m_0^2 m_c^2}{4M^4}\right) ~,
\label{fDs}
\ee
where $m_0^2= \langle\bar q \sigma_{\alpha\beta}
G^{\alpha\beta}q \rangle/ \langle \bar q q \rangle $
is a conventional parametrization
for the quark-gluon condensate. In the above,
we have omitted numerically insignificant
contributions of the gluon and four-quark condensates.
For consistency,
we do not take into account perturbative $O(\alpha_s)$ corrections to these
sum rules, since they are also not included
in the sum rule (\ref{fin}).


\section{Numerical analysis}

The principal nonperturbative input in the sum rule (\ref{fin})
are pion wave functions on the light-cone.
In ref. \cite{exclusive} a theoretical framework has been developed
to study these functions. In particular, it has been shown that
the wave functions can
be expanded in terms of matrix elements
of conformal operators which in leading logarithmic approximation
do not mix under renormalization.
For example, for the leading twist pion wave function one
finds an expansion in Gegenbauer polynomials,
\begin{equation}
\vp_\pi(u,\mu) = 6 u(1-u)\Big[1+a_2(\mu)C^{3/2}_2(2u-1)+
 a_4(\mu)C^{3/2}_4(2u-1)+\ldots\Big]\,,
\label{expansion}
\end{equation}
where all the nonperturbative information is included in the set
of multiplicatively renormalizable coefficients $a_n$, $n=2,4,\ldots$.
The corresponding anomalous dimensions are such that
the coefficients $a_n$ vanish for
$\mu \rightarrow \infty $, and the wave function is uniquely
determined by the first term in the expansion. Therefore,
this term is called asymptotic wave function.
Similar expansions also exist for the wave functions of nonleading
twist \cite{BF2}.

For practical applications it is important that the expansion in
conformal spin converges sufficiently fast.  How fast the wave functions
approach their well-known asymptotic form is still under debate.
However, there are indications \cite{radnew}
that the nonasymptotic deviations have been
overestimated previously. Corrections  to the asymptotic expressions
in next-to-leading (and in some cases also next-to-next-to-leading
order) in conformal spin are known for all wave functions which appear in
the sum rule (\ref{fin}). For details, we must refer the reader to the
original literature \cite{CZreport,BF2}.

In our numerical analysis we use the set of
nonleading twist wave functions proposed in ref. \cite{BF2}.
The explicit expressions and the values of the various parameters
are collected in Appendix~A.
Furthermore, we take
$f_\pi =132$ MeV,  $\mu_\pi(1 GeV)=1.65$, corresponding to
$ \langle \bar q q \rangle = -(243$~MeV$)^3 $,
$m_0^2=0.8$ GeV$^2$
and, in the charmed meson  channels,
$m_c=1.3$ GeV, $s_0=6$ GeV$^2$, $m_D=1.87$ GeV and $m_{D^\ast}=2.01$ GeV.
The same parameters  are also used to determine the decay constants
$f_D$ and $f_{D^{\ast}}$ from the sum rules (\ref{fD}) and (\ref{fDs}).
One obtains
\be
f_D = 170 \pm 10 MeV, ~~f_{D^*} = 240 \pm 20 MeV~.
\label{fDDstar}
\ee
The uncertainty quoted characterizes the variation with the
Borel parameter $M^2$ in the interval 1 GeV$^2< M^2 < $ 2 GeV$^2$.

Having fixed the input parameters,
one must find the range of values of $M^2$ for which the sum rule
(\ref{fin}) is reliable. The lowest possible  value of $M^2$ is
usually determined by the requirement that the terms
proportional to the highest inverse power of the Borel parameter
stay reasonably small. The upper limit is determined by
demanding that the continuum contribution does not get too large.
In the $D^*D\pi$ sum rule we take the interval
2 GeV$^2< M^2<$ 4 GeV$^2$.
 In this interval the twist 4 term
proportional to $M^{-2}$ does not exceed
$5\%$. Simultaneously, higher states contribute less than $30\%$.
The dependence of the r.h.s. of
eq. (\ref{fin}) on the Borel parameter is shown in Fig.~2a. As can be seen,
in the fiducial range of $M^2$ given above, the sum rule
is quite stable. From Fig. 2a one can directly read off the prediction
\be
f_Df_{D^{*}}g_{D^*D\pi }~= ~0.51 \pm 0.05~ GeV^2  ~.
\label{combinD}
\ee
Dividing this product of couplings
by the decay constants  (\ref{fDDstar}) finally yields
for the $D^*D\pi$ coupling constant:
\bq
g_{D^*D\pi}= 12.5\pm 1.0~,
\label{constD*Dpi}
\eq
where the error is understood to indicate the range of values
corresponding to the correlated variation of the results (\ref{fDDstar})
and (\ref{combinD})
within the fiducial intervals.
While the combination of couplings (\ref{combinD}) is not affected by
uncertainties
in the decay constants, it is more sensitive to the charm quark
mass and threshold $s_0$ than the coupling $g_{D^*D\pi}$ itself.

A few comments are in order.
The twist 3 terms contribute to the sum rule  at the level
of (50 $\div$ 60) $\%$ and are therefore as important as the
twist 2 contribution. On the other hand, the impact of twist 4 is small
amounting to about 5\%.
Two sources of uncertainties not included in eq. (\ref{combinD})
are the  nonasymptotic corrections to
the leading twist wave function  $\vp_\pi$ and
to the three-particle (twist 3) wave function $\vp_{3\pi}$.
The latter in turn induce
modifications in the two-particle (twist 3) wave functions $\vp_p$
and $\vp_\sigma$, apart from the corrections generated by the asymptotic
$\vp_{3\pi}$ wave function. In order to estimate the sensitivity
of our results to the nonasymptotic effects in $\vp_\pi$ and
$\vp_{3\pi}$ we drop these corrections altogether and recalculate
the product $f_Df_{D^{*}}g_{D^*D\pi }$. Remarkably,
the result changes by only  5$\%$. One can thus be confident that
the total uncertainty in eq. (\ref{combinD})
does not exceed 20$\%$.

We should emphasize that the above prediction can be directly
tested experimentally in the decay $D^* \rightarrow D \pi $.
With the value of $g_{D^*D \pi }$ given in eq. (\ref{constD*Dpi})
one predicts the decay width
\be
\Gamma( D^{*+} \rightarrow D^0 \pi^+)~ =
{}~ \frac{g_{D^*D\pi}^2}{24\pi m_{D^*}^2}|~\vec q ~|^3
{}~ = ~32 \pm 5 \,\mbox{\rm keV}  ~.
\label{Gamma}
\ee
Predictions for other charge combinations are
easily obtained from eq. (\ref{Gamma}) taking into account
the isospin relations (\ref{gd}) as well as small
differences in the phase space volumes:
\be
\Gamma( D^{*+} \rightarrow D^0 \pi^+)~ =
{}~2\cdot 1.1\,\Gamma (D^{*+} \rightarrow D^+ \pi^0)=~
2\cdot 0.72\,\Gamma (D^{*0} \rightarrow D^{0} \pi^0)~.
\label{widths}
\ee
The current experimental upper limit
\be
\Gamma( D^{*+} \rightarrow D^0 \pi^+)~ < ~ 89 ~\mbox{\rm keV}
\label{exp}
\ee
is obtained by combining the limit
$ \Gamma_{tot}(D^{*+}) < 131 ~ \mbox{\rm keV}$  \cite{ACCMOR}
with the branching ratio \\
$BR(D^{*+} \rightarrow D^0 \pi^+ ) = (68.1 \pm 1.0 \pm 1.3) \% $
\cite{CLEO}.
Our prediction is well below this upper limit.

The sum rule for $g_{D^*D\pi}$ given in eq. (\ref{fin}) is easily converted
into a sum rule for
the coupling $g_{B^*B\pi}= g_{\bar B^{*0}B^-\pi^+}$
by replacing  $c$ with  $b$, $D$ with  $\bar B$, and $D^*$ with $\bar B^*$.
The corresponding parameters are
$m_B=5.279$ GeV, $m_{B^*}=5.325$ GeV,
$m_b=4.7$ GeV, and $s_0=35$ GeV$^2$.
In addition, one has to evolute
the wave function parameters to a higher scale $\mu_b$ (see
Appendix A).
With these changes the
two-point sum rules (\ref{fD}) and (\ref{fDs}) yield
\begin{equation}
  f_B=140\,\mbox{\rm MeV}\,,~~~~f_{B^*}=160\,\mbox{\rm MeV}
\label{fB}
\end{equation}
with negligible uncertainties due to variation of $M^2$.
Using the same criteria as for the  $g_{D^*D\pi }$
sum rule the fiducial range in $M^2$ turns out to be
6 GeV$^2 < M^2 <~ $12 GeV$^2$. The stability of the sum rule for the
product of couplings $f_Bf_{B^*}g_{B^*B\pi }$ is illustrated in Fig.~2b.
We obtain
\be
f_Bf_{B^{*}}g_{B^*B\pi }=  0.64 \pm 0.06\,\mbox{\rm GeV}^2 ~,
\label{combinB}
\ee
and with eq. (\ref{fB})
\bq
g_{B^*B\pi}=29\pm 3~.
\label{41}
\eq
The hierarchy of various twists as well as the
uncertainty due to the nonasymptotic corrections are
found to be similar as in the case of $g_{D^*D\pi}$ .
Contrary to the latter, the coupling constant $g_{B^*B\pi}$
cannot be measured directly,
since the corresponding decay $B^* \rightarrow B \pi $ is kinematically
forbidden. However, the $B^*B\pi$ on-shell vertex is of a great
importance for understanding of the behaviour of heavy-to-light
form factors as will be discussed in Sect. 6.


\section{ Sum  rules from the short-distance expansion}

With the results of Sect. 4 at hand we are now also in a position
to study in more detail the soft-pion limit (\ref{SR1})
of the sum rule (\ref{fin}) which is obtained from the
correlation function (\ref{19})
at $p^2=(p+q)^2$ or, equivalently, for $ q\ra 0$.
As discussed in Sect. 2, in this limit one can apply a short-distance
expansion in terms of local operators with increasing
dimensions in contrast to the light-cone expansion involving
nonlocal operators with increasing twist.
In technical terms, at $q \rightarrow 0 $ only the lowest moments of
the wave functions contribute. Thus the integrals reduce to  overall
normalization factors.
The explicit expression for the invariant function $F$ in this limit
is directly obtained from eqs. (\ref{28}) and (\ref{35}):
\be
F(p^2)=\frac{m_cf_\pi}{m_c^2-p^2}\left[1+\frac{\mu_\pi}{3m_c}
\left(2+\frac{m_c^2}{m_c^2-p^2}\right)
+ \frac{10 \delta^2}{9 (m_c^2-p^2)}
\left( 1-\frac{m_c^2}{m_c^2-p^2}\right) \right]~,
\label{46}
\ee
where the parameter $\delta^2$ is specified
in Appendix A.
After Borel transformation in $p^2$ one has
\begin{eqnarray}
F(M^2)&= &m_cf_\pi e^{-\frac{m_c^2}{M^2}}\left[
1+\frac{2\mu_\pi}{3m_c}
+\frac1{M^2}\left( \frac{\mu_\pi m_c}3+\frac{10}9\delta^2
\right)-\frac{5m_c^2\delta^2}{9M^4}\right] ~.
\label{47}
\end{eqnarray}

As indicated in eq. (\ref{SR1}), the price for simplifying the
QCD representation of the correlation
function is a more complicated hadronic representation.
This in turn makes it more difficult to extract the ground state
contribution containing the $D^*D\pi$ coupling.
For illustration, we consider the contribution in the
dispersion representation of the correlation function (\ref{19}) from
the transition
of a given excited state in the $D^*$-channel with mass
$m_* > m_{D^*}$ to the ground state $D$-meson at $q \rightarrow 0$.
This contribution is proportional to
\be
\frac{1}{(m^2_*-p^2)(m_D^2-p^2)}~,
\label{mmm}
\ee
and, after Borel transformation, to
\be
\frac{1}{m^2_*-m_D^2}\left( e^{-\frac{m_D^2}{M^2}}-
e^{-\frac{m_*^2}{M^2}}\right)~.
\label{unsuppr}
\ee
Similar expressions hold for the ground state transition
$D^*\rightarrow D$  with $m_* = m_{D^*}$.
In the limit $m_D = m_{D^*}$, one has a double pole instead of
eq. (\ref{mmm}) and
\bq
\frac{1}{M^2}e^{-\frac{m_D^2}{M^2}}
\label{e2}
\eq
instead of eq. (\ref{unsuppr}). Clearly, the contribution (\ref{unsuppr})
is not exponentially suppressed relative to the ground state contribution
(\ref{e2}), and can therefore not be subtracted assuming duality.
On the other hand,
transitions involving excited states in both the initial
and the final states are suppressed by
Borel transformation with respect to the ground state transitions
and cause no problems.
Schematically, the complete hadronic part of the
invariant amplitude $F(M^2)$ can be written as follows:
\bq
F(M^2) \simeq \frac1{M^2}\left \{ \frac{m_D^2m_{D^*}f_Df_{D^*}}{m_c}
g_{D^*D\pi} +AM^2\right \} e^{-\frac{m_{D^*}^2+m_D^2}{2M^2}}
+ C\cdot e^{-\frac{m_*^2}{M^2}}~,
\label{44}
\eq
where the constant
$A$ incorporates all unsuppressed contributions of the type
 (\ref{unsuppr}), while the term proportional to $C$ contains all
exponentially suppressed contributions.

To get rid of the contaminating term
$ AM^2$ we follow ref. \cite{ioffe} and apply the
operator
\be
\left( 1-M^2\frac{d}{dM^2} \right)M^2e^{\frac{m_{D^*}^2+m_D^2}{2M^2}}
\label{oper}
\ee
to both representations (\ref{47}) and (\ref{44}).
Now it is possible to subtract the continuum contribution
by the same substitution (\ref{39})
which we have already employed to get the light-cone sum rule (\ref{fin}).
In this way we obtain a new sum rule for $g_{D^*D\pi}$:
\bq
f_Df_{D^*}g_{D^*D\pi}&=&\frac{m_c^2f_\pi}{m_D^2m_{D^*}}\left(
1-M^2\frac{d}{dM^2}\right)e^{\frac{m_D^2+m_{D^*}^2-2m_c^2}{2M^2}}
\nonumber
\\
&\times &\left[ M^2(1-e^{-\frac{s_0-m_c^2}{M^2}})
\left( 1+\frac{2\mu_\pi}{3m_c}\right)+
\frac{\mu_\pi m_c}{3}
 + \frac{10}{9}\delta^2-\frac{5m_c^2\delta^2}{9M^2}\right]~.
\label{48}
\eq
For the same input parameters and the range of $M^2$
leading to the prediction (\ref{constD*Dpi})
this sum rule yields
\bq
g_{D^*D\pi}=11\pm 2~.
\label{49}
\eq
Similarly, replacing in eq. (\ref{48}) the charmed meson parameters by the
corresponding quantities in the beauty channel, one finds
\bq
g_{B^*B\pi}=28\pm 6 ~.
\label{50}
\eq
As compared with the predictions (\ref{constD*Dpi}) and
(\ref{41})
the uncertainties are larger by a factor of two due to the
worse stability of the sum rule (\ref{48}) against variation of $M^2$.
The agreement of the results indicates selfconsistency
of the sum rule approach and gives support to the approximations used
in the pion wave functions.

Furthermore, one can show that the sum rule  (\ref{48})
is actually equivalent to
the sum rule obtained in ref. \cite{EK85} using  external field
techniques.
Indeed, applying the usual reduction formalism to the pion, one can
rewrite  the correlation function (\ref{19}) in the following form:
\bq
F_\mu(p,q)=(q^2-m_\pi^2)\int d^4x~ d^4y~ e^{i(px-qy)}
\langle 0\mid T\{\bar{d}(x)\g_\mu c(x),
\phi_\pi(y),\bar{c}(0)\g_5u(0)\}\mid 0\rangle ~,
\label{q1}
\eq
where $\phi_\pi(y)$ is the interpolating  pion field.
According to PCAC
\bq
\phi_\pi(y)=\frac{\partial^\mu j_\mu^5(y)}{f_\pi m_\pi^2}~,
\label{q2}
\eq
$ j_\mu^5(y)=\bar{u}(y)\g_\mu\g_5d(y)$ being the axial current.
Subsituting eq. (\ref{q2}) in eq. (\ref{q1}) and
integrating  by parts, one gets \footnote{
In addition, one obtains two-point correlation functions because of
contact terms. These do not lead to  double poles in $p^2$
in the relevant dispersion relation and are therefore
eliminated by applying the  differentiation operator (\ref{oper}).}
\bq
F_\mu(p,q)
&=&i\frac{q^2-m_\pi^2}{f_\pi m_\pi^2}T_{\mu\rho}(p,q)q^\rho
= -\frac{i}{f_\pi}T_{\mu\rho}(p,q)q^\rho~,
\label{Fmu}
\eq
with
\bq
T_{\mu\rho}(p,q)&=&\int d^4x~d^4y~e^{i(px-qy)}
\langle 0|T\{ \bar{d}(x)\g_\mu c(x), j_\rho^5(y),
\bar{c}(0)\g_5u(0)\}|0\rangle~.
\label{q3}
\eq

Instead of dealing with the three-point correlation
function $T_{\mu\rho}(p,q)$ directly, it is more convenient
to consider
the following two-point correlation function in the constant external
axial field  $A_\mu^5$:
\begin{equation}
T^A_{\mu}(p,q)=\int d^4x~e^{ipx}
\langle0 \mid T\{ \bar{d}(x)\g_\mu c(x),\bar{c}(0)
\g_5u(0) \}\mid  0\rangle_A ~.
\label{q4}
\end{equation}
It is assumed that a term $A^{5\mu}j_\mu^5$ corresponding to the interaction
of the external field with the light quarks is  added to the QCD Lagrangian.
To first order in the external field, this correlation function is
given by $$ T^A_{\mu}(p,q) = T_{\mu\rho}(p,q)A^{5\rho}\, ,$$ with
$T_{\mu\rho}(p,q)$ as defined in eq. (\ref{q3}).
In the above sense, the two-point correlation function (\ref{q4})
in the constant axial field, is
equivalent to the three-point function (\ref{q3}) and, via
the PCAC relation (\ref{q2}),
also to the two-point correlation function (\ref{19})  at $q \ra 0$
\footnote{ As a side remark, the
light-cone approach leading to eq. (\ref{fin}) corresponds
to a calculation in the background
of a {\em variable}  external axial field \cite{BBK,BF1}.}.
Consequently, the sum rules obtained in refs. \cite{EK85} and
\cite{Cetc94} should coincide  with each other
and with the sum rule (\ref{48}) derived in this paper.

In particular, the expression (\ref{47})
for $F(M^2)$ can be compared with the result
given in eq. (19) of ref. \cite{EK85} and
in eq. (2.15) of ref. \cite{Cetc94},
after normalization and kinematical structures are adjusted properly.
In refs. \cite{EK85,Cetc94}
the correlation function (\ref{19}) is separated as follows:
\be
F_\mu = Aq_\mu + B(2p_\mu +q_\mu) ~,
\label{choice}
\ee
and the sum rule is obtained by evaluating of the invariant function $A$
for $q \ra 0$.
In terms of the invariant amplitudes defined in eq. (\ref{220})
one has
\be
A= F-\frac{\tilde F}2, ~ B= \frac{\tilde F }2~.
\label{our}
\ee
Hence, for comparison we need also the second invariant
function $\tilde F$, which we have not discussed so far,
but which can be calculated along the same lines.
Adding this contribution to eq. (\ref{47}), we have checked that
our result for the invariant amplitude $A$ coincides
with the one presented in ref. \cite{Cetc94},
apart from terms proportional to $m_0^2$ which,
being associated with twist 5 contributions
in the light-cone sum rules,
are neglected in our approximation. Numerically, this terms are
not important.
On the other hand, we disagree with ref. \cite{EK85} in the non-leading
terms proportional to $\delta^2$.

Although it is legitimate to use different Lorentz decompositions
of the correlation function (\ref{19}) in order to derive the
desired sum rule,
we think that the choice adopted in the present paper is more adequate
for the following reason. Since the vector current $\bar q \g_\mu c $
is not conserved, it not only couples to $J^P=1^-$ vector mesons,
but also  to $J^P=0^+$ scalar mesons ($D_0$). The corresponding transition
matrix element is proportional to the momentum $p_\mu$:
\be
\langle 0 \mid \bar q \g_\mu c \mid D_0\rangle = f_{D_0}m_{D_0}p_\mu~.
\label{scalar}
\ee
The mass of the ground state $D_0$ meson is
expected to be in the vicinity of 2400 MeV
which is not far from the mass of the $D^*$ and below the accepted
continuum threshold in the $D^\ast$ channel.
For this reason,
the $D_0$ contribution should be added to the sum rule.
Unfortunately, this introduces additional uncertainties
in the hadronic representation as is the case, for example,
in ref. \cite{Cetc94}.
In contrast, our sum rules based on the invariant function
$F$ defined in eq. (\ref{220})
are not affected by scalar meson contributions, which is a clear
advantage.

A calculation rather similar to ref. \cite{Cetc94}, but with particular
emphasis on the heavy quark limit, has been carried out
earlier in ref. \cite{ovch}. Very recently, another
calculation in the heavy quark limit
appeared in ref. \cite{GY94} using the external field technique.
Unfortunately, in this paper
a wrong expression for the induced quark condensate  in
the external field is used, as can be seen by consulting
refs. \cite{EK85,ga}.
The error can be traced back to the equation
of motion for the quark field in presence of the external field which
is modified from $i\!\not\!\!\!D q = 0 $ to
$i\!\not\!\!\!D q = \not\!\!\!A^5 q $.
By this modification the axial current insertions into the vacuum quark
legs are properly taken into account.
The numerical comparison of these different calculations is left for the
concluding section.

\section{Pole model for $D\ra\pi$ and
$B\ra\pi$ form factors and QCD sum rules}

The couplings $g_{D^*D\pi}$ and $g_{B^*B\pi}$ fix the normalization
of the form factors of the heavy-to-light transitions $D\ra\pi$ and
$B\ra\pi$, respectively, in the pole model description [5,6]. This
model is based on the vector dominance idea suggesting a momentum
dependence dominated by the $D^*$ and $B^*$ poles, respectively. More
definitely, the form factor $f^+_D(p^2)$ defined by the matrix element
\be
\langle \pi(q) \mid \bar d\g_\mu c \mid D(p+q) \rangle =
2f^+_D(p^2)q_\mu + ( f^+_D(p^2) -f^-_D(p^2))p_\mu
\label{formfdef}
\ee
is predicted to be given by
\be
f^+_D(p^2)= \frac{f_{D^*}g_{D^*D\pi}}{2m_{D^*}(1-p^2/m_{D^*}^2)}~.
\label{onepole}
\ee
An analogous expression holds for the form factor $f^+_B(p^2)$.

It is difficult to justify the pole model from first principles.
Generally, it is believed that the vector dominance approximation is valid
at zero recoil, that is at $p^2\ra m_D^2$. Arguments based
on heavy quark symmetry suggest a somewhat larger region
of validity characterized by
$(m_D^2-p^2)/m_c \sim O(1 $GeV). However,  there are no convincing arguments
in favour of this model to be valid also at small values of $p^2$ which are
most interesting from a practical point of view. Therefore, the finding
\cite{BBD1,Ball} that
the pole behaviour is consistent with the $p^2$ dependence at $p^2\ra 0$
predicted by sum rules, is very remarkable. Meanwhile, this
claim has been confirmed by independent calculations within the
framework of the light-cone sum rules \cite{BKR}.

In this section we want to demonstrate that not only the shape but
also the absolute normalization of the above form factors appears to be
comparable with the pole model description. This assertion is non-trivial,
since contributions of several low-lying resonances in the $D^*$ or
$B^*$ channel
could still mimic the $p^2$ dependence of a single pole, but the relation
to the coupling $g_{D^*D\pi}$ or $g_{B^*B\pi}$ should then
be lost \cite{IW90}. However, despite of the overall agreement in the
mass range of $D$ and $B$ mesons, there is a clear
disagreement on the
asymptotic dependence of the form factors on the heavy mass.
The QCD sum rules on the light-cone provide a unique framework to
examine these issues, since both the form factors $f^+_{D,B}(p^2)$
at $m_{c,b}^2-p^2\geq O(1$ GeV$^2$) and the couplings
$g_{D^*D\pi}$ and $g_{B^*B\pi}$ can be
calculated from the same correlation function (\ref{19})
using the same technique.
In addition, contrary to conventional sum rules \cite{BBD1},
this approach leads to consistent results in the heavy quark limit
\cite{ABS}.

The detailed derivation of the light-cone sum rules for the
$D\ra\pi$ and $B \ra \pi $ form factors is discussed in ref. \cite{BKR}
(see also
refs. [19-21]). Here we just mention that
the sum rule for $f^+_D(p^2)$ is obtained by matching
the expressions (\ref{28})   and (\ref{35})
for the invariant amplitude $F(p^2, (p+q)^2)$
in terms of the pion wave functions
with the hadronic representation
\be
F(p^2, (p+q)^2) = \frac{2m_D^2f_Df^+_D(p^2)}{m_c(m_D^2-(p+q)^2)}
+ \int_{s_0}^{\infty} \frac{\rho^h(p^2,s)ds}{s- (p+q)^2} ~.
\label{physpart}
\ee
In the above, the pole term is due to the ground state in the heavy
channel, while the excited and continuum states
are taken into account by the dispersion integral above the threshold
$s_0$. Invoking duality, the latter contributions are cancelled against the
corresponding pieces in eqs. (\ref{28}) and (\ref{35}). After Borel
transformation
in the variable $(p+q)^2$, the resulting sum rule takes the form
\bq
\lefteqn{f^+_D( p^2)= \frac{f_\pi m_c^2}{2f_Dm_D^2}
\Bigg \{\int_\Delta^1\frac{du}{u}
\exp\left[\frac{m_D^2}{M^2}-\frac{m_c^2-p^2(1-u)}{uM^2}\right]
\Phi_2(u,M^2,p^2)}
\\&&{}
-\int_0^1\!\!u du\!\int \frac{{\cal D}\alpha_i
\Theta( \alpha_1+u\alpha_3-\Delta)}{(\alpha_1+u\alpha_3)^2}
\exp\!\left[\frac{m_D^2}{M^2}-\frac{m_c^2-p^2
(1-\alpha_1-u\alpha_3)}{(\alpha_1+
u\alpha_3)M^2}\right]\!\Phi_3(u,M^2,p^2) \Bigg\}\nonumber~,
\label{formSR}
\eq
where
\begin{eqnarray}
\Phi_2& = &\vp_\pi(u) +
\frac{\mu_\pi}{m_c}\Bigg[u \vp _{p}(u)
+ \frac16 \vp_{\sigma }(u)
\left(2 + \frac{m_c^2+p^2}{uM^2}\right)\Bigg]
\nonumber\\
&&{}
- \frac{4m_c^2g_1(u)}{u^2M^4} - \frac{2G_2(u)}{uM^2} \left(1+
\frac{m^2_c+p^2}{uM^2} \right) ~,
\label{XXX}
\\
\Phi_3&=& \frac{2f_{3\pi}}{f_{\pi}m_c}
\varphi_{3\pi}(\alpha_i)
\left[1-\frac{ m^2_c -p^2 }{(\alpha_1+u\alpha_3)M^2}\right]
\nonumber\\&&{}
-\frac1{uM^2} \Bigg[2\vp_\perp (\al_i)-\vp_\parallel (\al_i)+
2\tilde{\vp}_\perp (\al_i)-\tilde{\vp}_\parallel (\al_i)\Bigg]~,
\label{xxz}
\end{eqnarray}
and
$\Delta = (m_c^2-p^2)/(s_0-p^2) $.
The notation is as in eq. (\ref{fin}).
Improving the approximation given in ref. \cite{BKR} we have added
the contributions of three-particle wave functions of twist 4.
The analogous sum rule  for the  $B \ra \pi$ form factor follows from
the above by replacing $c \ra b $ and $D \ra \bar B $, and by
rescaling $\mu_\pi$ and the wave function parameters from
$\mu_c$ to $\mu_b$.

The maximum momentum transfer $p^2$ at which these sum rules are
applicable is estimated to be about
15 GeV$^2$ for B mesons, and 1 GeV$^2$ for D mesons.
For numerical evaluation we use the approximations of the wave functions
given in Appendix A.
We emphasize that the input here is exactly the same as in the
calculation of the couplings
$g_{D^*D\pi}$ and $g_{B^*B\pi}$ .
The form factor $f^+_D( p^2)$ resulting from the sum rule (\ref{formSR})
is plotted in Fig. 3a,
together with the corresponding prediction (\ref{onepole}) of the
pole model. We see that in the region of overlap
both calculations approximately agree with each other. To a lesser extent,
this also applies to the form factor $f^+_B( p^2)$ illustrated in
Fig. 3b \footnote{The dependence of eq. (\ref{formSR})
on the Borel parameter is weak \cite{BKR}. For definiteness,
we take here $M^2=4$ GeV$^2$ for the $D\ra\pi$ form factor
and $M^2=10$ GeV$^2$ for the $B\ra\pi$ form factor. }.
Quantitatively, at $p^2=0$ we find
\be\label{fD0}
f^+_D(0)_{SR}= 0.66,~~     f^+_D(0)_{PM}=0.75  ~,
\ee
and
\be\label{fB0}
f^+_B(0)_{SR}= 0.29,~~     f^+_B(0)_{PM}= 0.44 ~.
\ee

Thus, in the regions $m_Q^2-p^2 > O(1$ GeV$^2$) with $Q=c$ and $b$,
respectively, the numerical agreement between the light-cone sum rule
and the pole model is better than 15\% for  $f^+_D$, but only within
50\% for  $f^+_B$.

At this point, we must add a word of caution concerning the
applicability of the pole model too
far away from the zero recoil point, in particular at $p^2=0$.
The two descriptions differ markedly in the asymptotic dependence
of the form factors on the heavy mass. Focusing on $B$ mesons and
using the familiar scaling laws
\be
f_B\sqrt{m_B}= \hat{f}_B, ~~~~ f_{B^*}\sqrt{m_B} = \hat{f}_{B^*} ~,
\label{consthq}
\ee
and
\be\label{gscale}
g_{B^*B\pi} = \frac{2 m_B}{f_\pi}\cdot \hat g ~,
\ee
which are expected to be valid at $m_b\to\infty$ modulo logarithmic
corrections, one obtains
\be
f^+_B(0)_{PM} \sim 1/\sqrt{m_B} ~,
\ee
whereas the light-cone sum rule (79) yields
\cite{CZ90}
\be\label{SRlim}
f^+_B(0)_{SR} \sim 1/{m_B^{3/2}}~.
\ee
This result rests on the QCD prediction \cite{exclusive} of
 the behaviour of the leading twist pion
wave function near the end point, that is  $\vp_\pi(u)\sim 1-u$
at $u \ra 1$.
It should be noted that the contribution estimated by the sum rules
corresponds to the so-called Feynman mechanism.
 In the case of
heavy-to-light transitions it leads to
the same asymptotic behaviour as the
hard rescattering mechanism \cite{CZ90,BurDon}.
Recently it has been shown \cite{Akhoury}
that the power behaviour (\ref{SRlim})
of hard rescattering is not modified by the
Sudakov type double logarithmic corrections.
We believe that the light-cone sum rules reproduce
the true asymptotic behaviour, although a
rigorous proof in QCD is still lacking.
On the other hand, we see no theoretical justification for
extrapolating the pole model to the region $p^2=0$.
The solution suggested by Fig. 3 is to match
the two descriptions in the region of intermediate momentum transfer
$ p^2\simeq m_Q^2-O(1$GeV$^2)$.

Referring for a detailed discussion to refs. \cite{ABS} and \cite{X}
we want to
emphasize  that the light-cone sum rules seem to be generally
consistent with the heavy quark expansion. In particular,
the light-cone sum rule (\ref{fin}) correctly reproduces the heavy quark
mass dependence of the coupling $g_{B^*B\pi}$. Fitting our
predictions for $g_{B^*B\pi}$ and $g_{D^*D\pi}$ to the form
\be\label{1/m}
g_{B^*B\pi} = \frac{2 m_B}{f_\pi}\cdot \hat g
\Bigg[1+\frac{\Delta}{m_B}\Bigg]
\ee
and the analogous expression for $g_{D^*D\pi}$,
we find for the coupling $\hat g$ and the strength $\Delta$
of the $1/m_Q$ correction:
\be
\hat g =0.32\pm 0.02~,~~ \Delta =(0.7 \pm 0.1)~ GeV ~.
\label{fit}
\ee

Furthermore, we are able to make a numerical prediction for the
theoretically interesting ratio
\be
\frac{g_{B^*B \pi}f_{B^*}\sqrt{m_{D}}}
{g_{D^*D \pi}f_{D^*}\sqrt{m_{B}}} \simeq ~0.92 ~.
\label{ratio3}
\ee
This ratio is unity in the heavy quark limit and is shown to be
subject to $1/m_Q$ corrections only in the next-to-leading order
\cite{BLN}. Our result (\ref{ratio3}) is nicely consistent with this
expectation. The deviation from unity also agrees in magnitude with
the estimate in ref. \cite{Cetc94}, but has a different sign. This is due to
a sizeable difference in the ratio $f_{B^*}/f_{D^*}$. While the values
of the decay constants
given in eqs. (\ref{fDs}) and (\ref{fB}) yield
\be
\frac{f_{B^*}\sqrt{m_{B}}}{f_{D^*}\sqrt{m_{D}}}=1.12
\label{rrr}
\ee
in agreement with the expectation quoted in ref. \cite{BLN},
the latter ratio
turns out to be larger by 30\% if calculated from $f_{B^*}$ and $f_{D^*}$
as assumed in ref. \cite{Cetc94}.


\section{Summary and conclusions}

We have presented a comprehensive analysis of the
pion couplings to heavy mesons in the framework of QCD sum rules.
The main new result of this paper is the light-cone sum rule (\ref{fin})
providing the numerical estimates for $g_{D^*D\pi}$ and
$g_{B^*B\pi}$ given in
eqs. (\ref{constD*Dpi}) and (\ref{41}), respectively. The
decay width  $ \Gamma(D^* \ra D\pi)$ predicted in eq. (\ref{Gamma}) turns
out to be three times smaller than the present experimental upper
limit. We have compared our results to earlier QCD sum rule
calculations [10-13], and resolved the existing discrepancies.

A rather complete compilation of estimates
\footnote{We have not included the
results of ref. \cite{EK85} since to our knowledge this analysis is
being reconsidered \cite{PC:Eletsky}. The result of ref. \cite{GY94}
is omitted for reasons explained in Sect. 5.
} on the pion couplings
to heavy mesons is given in Table 1. In the first row we show
predictions on the reduced coupling $\hat g$
defined in eq. (\ref{gscale}). As one can see, the values
obtained by combining the nonrelativistic constituent quark model with PCAC
\cite{IW90,NW,Yan}
are roughly two times larger than the values favoured by our sum rule.
However, more recent analyses \cite{CG,Ametal} combining chiral HQET
with experimental
constraints on $D^*$ decays tend to give somewhat
smaller values of $\hat g$. Moreover, another recent calculation
\cite{BardeenHill} based on the extended Nambu-Jona-Lasinio model and
chiral HQET is in perfect agreement with our estimate.

The next two rows list the estimates of the couplings $g_{B^*B\pi}$
and $g_{D^*D\pi}$. These predictions are even wider spread. Quark models
\cite{Eichtetal,DoXu} seem to give the strongest couplings, whereas
SU(4) symmetry \cite{Kam} and the reggeon quark-gluon string model \cite{KN}
predict a relatively small coupling. Two comments are in order concerning
the analysis of ref. \cite{Cetc94}. Firstly, these
predictions are consistently lower
than ours. There are several reasons for that: the different Lorentz
decompositions (\ref{220}) and (\ref{choice})
of the correlation function (\ref{19}),
the differences between the sum rule (\ref{fin}) and the soft-pion limit
(\ref{48}) of it, the different regions of the Borel parameter $M^2$, and the
different values used for the decay constants $f_{D^{(*)}}$ and
$f_{B^{(*)}}$.
In fact, as can be seen in Fig. 2,
the couplings shrink with $M^2$. However, given the reliability criteria,
generally accepted for sum rules, we see no possibility to shift $M^2$
to larger values beyond the regions considered in this paper,
in contrast to ref. \cite{Cetc94}.
Secondly,
we find it inconsistent to include the perturbative gluon correction in
the estimates of $f_{D,D^*}$
and $f_{B,B^*}$, since they are not included in the
sum rule for the combination of couplings $f_Df_{D^*}g_{D^*D\pi}$ and
$f_Bf_{B^*}g_{B^*B\pi}$. At least, we see no convincing argument in favour
of such a procedure. For these two reasons we believe that the couplings are
underestimated in ref. \cite{Cetc94}.

For convenience and direct comparison with future measurements the decay width
\\$\Gamma(D^{*+}\ra D^0 \pi^+)$ as calculated from $g_{D^*D\pi}$ or $\hat g$
is quoted in the last row of Table 1. The widths in the channels
$D^{*+} \rightarrow D^+ \pi^0$ and $D^{*0} \rightarrow D^{0} \pi^0$ are
related to the above by coefficients which can be read off from
eq. (\ref{widths}).
Note that in contrast to the evaluation of $\Gamma(D^*\ra D\pi)$ from
$g_{D^*D\pi}$ in this paper and in ref. \cite{Cetc94} the estimates
in refs. \cite{Yan,CG} using the reduced coupling $\hat g$ do not include
$1/m$ corrections. However, the latter are important as can be seen from
eq. (\ref{fit}).

In addition, we have examined the pole model for the $B\to\pi$ and
$D\to\pi$ form factors. Using our results on the $g_{B^*B\pi}$
and $g_{D^*D\pi}$ coupling constants, we
have found approximate numerical agreement between the pole model description
and the direct sum rule calculation. However, the dependence on the
heavy quark mass is found to be completely different in the region of small
momentum transfers.
We have argued in favour of the sum rule approach. Moreover,
writing a heavy quark
expansion for the couplings $g_{B^*B\pi}$ and $g_{D^*D\pi}$ we have determined
the expansion coefficients, in particular, the leading $1/m$ correction.

Last but not least, we have discussed in some detail the theoretical
foundations and advantages of the light-cone sum rules,
complementing the work of refs. [14-19].
We believe that this approach is especially suitable for the study
of heavy-to-light decay form factors, and coupling constants of
the type considered in this paper. Further obvious applications include
the radiative decays $D^*\to D\gamma$ and $B^*\to B\gamma$.
Since the photon wave functions
are expected to deviate less from their asymptotic forms
than the pion wave functions \cite{BBK}, these decays
should provide a rather conclusive consistency
check of the light-cone approach.

\bigskip
\bigskip
\noindent
{\Large \bf Acknowledgements}

\bigskip
V.M. Belyaev is grateful to DAAD for financial support during his
visit at the University of Munich. This work is also partially supported
by the EC grant  INTAS-83-283.

\section*{Appendix A}
\app
For convenience, we collect here the explicit
expressions for the pion wave functions used in our numerical
calculations and specify the values of the parameters involved.

For the leading twist two wave function we take \cite{BF1}
\begin{equation}
\vp_\pi(u,\mu) = 6 u(1-u)\Big[1+a_2(\mu)C^{3/2}_2(2u-1)+
 a_4(\mu)C^{3/2}_4(2u-1)\Big]
\label{expansion1}
\end{equation}
with the Gegenbauer polynomials
\bq
C_2^{3/2}(2u-1)=\frac{3}2[5(2u-1)^2-1]~,
\nonumber
\\
C_4^{3/2}(2u-1)=\frac{15}8[21(2u-1)^4-14(2u-1)^2+1]~,
\label{G2}
\eq
and the coefficients  $a_2=\frac23$, $a_4=0.43$
corresponding to the normalization point $\mu=0.5$ GeV.
In the present applications the appropriate scales are set by the
typical virtuality of the heavy quark. We choose
\bq
\mu_c = \sqrt{m_D^2-m_c^2} \simeq 1.3\,\mbox{\rm GeV}\,,~~~
\mu_b =\sqrt{m_B^2-m_b^2} \simeq 2.4\, \mbox{\rm GeV}\,.
\label{scales}
\eq
Renormalization group evolution of the coefficients $a_2$ and
$a_4$ to these higher scales yields
\bq
a_2(\mu_c) = 0.41 ,~ a_4(\mu_c) = 0.23~,
\label{a24c}
\nonumber
\\
a_2(\mu_b) = 0.35 ,~ a_4(\mu_b) = 0.18 ~.
\label{a24b}
\eq
We stress that the value of $\vp_\pi$ at $u=1/2 $  varies by only 2\%
when the scale $\mu$ is increased from 0.5 GeV to 2.4 GeV. Obviously,
one can neglect this effect given the 15\% uncertainty in the value
of $\vp_\pi(u=1/2,\mu=0.5$ GeV$)$
quoted in eq. (\ref{WF12}).

According to the analysis in refs. \cite{BF1,BF2} the set of wave
functions of twist three is uniquely specified by the choice of the
three-particle wave function $\vp_{3\pi}$. Taking into
account the contributions to $\vp_{3\pi}$ up to
next-to-next-to-leading order in conformal spin, one has
\bq
\vp_{3\pi}(\al_i)&=&360 \al_1\al_2\al_3^2
\Big(1+\omega_{1,0}\frac12(7\al_3-3)
\nonumber
\\
&&{}+\omega_{2,0}(2-4\al_1\al_2-8\al_3+8\al_3^2)
+\omega_{1,1}(3\al_1\al_2-2\al_3+3\al_3^2)\Big] ~.
\label{3pi}
\eq
This implies for the
two-particle wave functions of twist three
\cite{BF2}:
\bq
\vp_p(u)&=&1+B_2\frac12(3(u-\bar{u})^2-1)+B_4\frac18(35(u-\bar{u})^4
-30(u-\bar{u})^2+3)
\label{bbb}
\eq
and
\bq
\vp_\sigma (u)&=&6u\bar{u}\left[ 1+C_2\frac{3}2(5(u-\bar{u})^2-1)
+C_4\frac{15}8(21(u-\bar{u})^4-14(u-\bar{u})^2+1)\right]~,
\label{tw3}
\eq
where
\begin{eqnarray}\label{wf3}
B_2&=&30R\,,
\nonumber\\
B_4&=&\frac32R(4\omega_{2,0}-\omega_{1,1}-2\omega_{1,0})\,,
\nonumber\\
C_2&=&R(5-\frac12\omega_{1,0})\,,
\nonumber\\
C_4&=&\frac1{10}R(4\omega_{2,0}-\omega_{1,1})\,,
\label{ccc}
\end{eqnarray}
with
\begin{eqnarray}
R&=&\frac{f_{3\pi}}{\mu_\pi f_\pi}\,.
\label{RR}
\end{eqnarray}
The coefficients $f_{3\pi}$ and
$\omega_{i,k}$ have been determined at the normalization
point $\mu = 1$ GeV from QCD sum rules \cite{CZreport}:
\bq
f_{3\pi} = 0.0035~\mbox{\rm GeV}^2,~~
\omega_{1,0}= -2.88\,,~~ \omega_{2,0}= 10.5\,,~~ \omega_{1,1}= 0~.
\eq
After renormalization \cite{BB} to the relevant scales (\ref{scales}),
we get
\bq
f_{3\pi}(\mu_c) = 0.0032~ \mbox{\rm GeV}^2,~~
\omega_{1,0}(\mu_c)= -2.63 ,~~
\omega_{2,0}(\mu_c)= 9.62 ,~~
\omega_{1,1}(\mu_c)= -1.05 ~,
\nonumber\\
f_{3\pi}(\mu_b) = 0.0026~ \mbox{\rm GeV}^2,~~
 \omega_{1,0}(\mu_b)= -2.18,~~
 \omega_{2,0}(\mu_b)= 8.12,~~
 \omega_{1,1}(\mu_b)= -2.59\,~.
\eq
The corresponding numerical values of the coefficients (\ref{ccc}) are
as follows:
\bq
B_2(\mu_c)=0.41,~~B_4(\mu_c)=0.925,~~C_2(\mu_c)=0.087,~~C_4(\mu_c)=0.054\,,
\nonumber\\
B_2(\mu_b)=0.29,~~B_4(\mu_b)=0.58,~~C_2(\mu_b)=0.059,~~C_4(\mu_b)=0.034\,.
\eq
In addition, the running of light-quark masses induces a scale dependence
of the parameter
$\mu_\pi= m_\pi^2/(m_u+m_d)$:
\be
\mu_\pi(1\, \mbox{\rm GeV}) = 1.65\,\mbox{\rm GeV},~~
\mu_\pi(\mu_c) = 1.76\,\mbox{\rm GeV},~~
\mu_\pi(\mu_b) = 2.02\,\mbox{\rm GeV}\,.
\ee

The wave functions of twist four are more numerous. The complete set
given in ref. \cite{BF2} (see also ref. \cite{gorsky}) includes
four three-particle
wave functions. However, in leading and
next-to-leading order in conformal spin, these are specified
by only two parameters:
\bq
\vp_\perp (\al_i)&=&30\delta^2 (\al_1-\al_2)\al_3^2[\frac13+2
\varepsilon (1-2\al_3)] ~,
\nonumber
\\
\vp_\parallel (\al_i)&=&120\delta^2\varepsilon (\al_1-\al_2)\al_1\al_2\al_3~,
\nonumber
\\
\tilde{\vp}_\perp (\al_i)&=&30\delta^2\al_3^2(1-\al_3)[\frac13+2
\varepsilon (1-2\al_3)] ~,
\nonumber
\\
\tilde{\vp}_\parallel (\al_i)&=&-120\delta^2\al_1\al_2\al_3[\frac13+
\varepsilon (1-3\al_3)] ~.
\label{tw4gluon}
\eq
The two-particle twist 4 wave functions are related to these
by equations of motion.
To the above order in conformal spin  they involve no new parameters and are
given by
\begin{eqnarray}
g_1(u)&=&\frac{5}2\delta^2\bar{u}^2u^2+\frac{1}{2}\varepsilon\delta^2[
\bar{u}u(2+13\bar{u}u)+10u^3\ln u(2-3u+\frac65u^2)
\nonumber
\\
&&{}+10\bar{u}^3\ln \bar{u}(2-3\bar{u}+\frac65\bar{u}^2)]\,,
\nonumber
\\
g_2(u)&=&\frac{10}3\delta^2\bar{u}u(u-\bar{u})\,,
\nonumber\\
G_2(u)&=& \frac53\delta^2 \bar u ^2 u^2 ~.
\label{tw4}
\end{eqnarray}
One of the parameters is defined by the matrix element
\be
\langle \pi |g_s\bar{d}\tilde{G}_{\al\mu}\gamma^\al u|0 \rangle=
i\delta^2f_\pi q_\mu ~.
\label{delta}
\ee
The QCD sum rule estimate of ref. \cite{nov} yields
$\delta^2=0.2$ GeV$^2$ at $\mu=1$ GeV. The remaining
parameter is associated with the
deviation of twist four wave functions from their
asymptotic form. At $\mu=1$ GeV it takes the value
$\varepsilon =0.5$  \cite{BF2}.
Renormalization to the relevant scales (\ref{scales})
gives
\bq
\delta^2(\mu_c)=0.19,~~ \varepsilon(\mu_c) =0.45\,,
\nonumber\\
\delta^2(\mu_b)=0.17,~~ \varepsilon(\mu_b) =0.36\,.
\eq
This completes the description of the pion wave functions,
as far as it is needed for the applications in this paper.

\appende

\section*{Appendix B}
\app
Here we derive the substitution (\ref{39}) used
in the sum rule (\ref{fin}) in order to subtract the continuum contribution.
For this purpose we have to write the
invariant amplitude
$F$ given by eqs. (\ref{28}) and (\ref{35})
in the form of a double dispersion integral:
\be
F( p^2, (p+q)^2)=\int^\infty_{m_c^2} \frac{ds_1}{s_1-p^2}
\int^\infty_{m_c^2} \frac{ds_2}{s_2-(p+q)^2}\rho^{QCD}(s_1,s_2)~.
\label{repr}
\ee
Focusing first on the leading contribution (\ref{Fzeroth}):
\be
F(p^2, (p+q)^2)= m_cf_\pi
\int^1_0 \frac{du ~\vp_\pi(u)}{m_c^2-(p+uq)^2} =
m_cf_\pi\int^1_0 \frac{du~\vp_\pi(u)}{m_c^2-(p+q)^2u-p^2(1-u)}~,
\label{integr}
\ee
and changing variable from $u$ to $(m_c^2-p^2)/(s-p^2) $,
one obtains
\be
F( p^2, (p+q)^2)=m_cf_\pi\int^\infty_{m_c^2} \frac{ds~
\vp_\pi(u(s))}{(s-(p+q)^2)(s-p^2)}  ~.
\label{integrs}
\ee
In general, the wave function $\vp_\pi(u)$ can be expressed as
a power series in $(1-u)$:
\be
\vp_\pi(u) = \sum_k a_k(1-u)^k =
\sum_ka_k\left(\frac {s-m_c^2}{s-p^2} \right)^k~.
\label{sum}
\ee
Substituting this representation into eq. (\ref{integrs})
and introducing formally two variables $s_1$ and $s_2$ instead of $s$,
it is easy to rewrite this expression in the form (\ref{repr})
with the double spectral density
\be
\rho^{QCD}(s_1,s_2)=m_cf_\pi\sum_k \frac{(-1)^ka_k}{\Gamma(k+1)}
(s_1-m_c^2)^k\delta^{(k)}(s_1-s_2)~.
\label{dens}
\ee
The validity of eq. (\ref{dens}) can easily be checked by direct calculation.
The above derivation may seem tricky. However, there is a
convenient general method \cite{rad2} to find the appropriate
double spectral densities. One takes the Borel transformed
invariant amplitude $F( M_1^2, M_2^2)$ and performs two more
Borel transformations in the  variables $\tau_1=1/M^2_1$ and
$\tau_2=1/M^2_2$, to get
\be
{\cal B}_{\sigma_1}{\cal B}_{\sigma_2}F(1/\tau_1,1/\tau_2) =
\rho^{QCD}(1/\sigma_1,1/\sigma_2) ~.
\ee
Details and useful examples can be found in ref. \cite{BB94}.

To proceed, we apply a  double Borel transformation to the
dispersion integral (\ref{repr}) with $\rho^{QCD}$ given by eq. (\ref{dens}):
\be
{\cal B}_{M_1^2}{\cal B}_{M_2^2}F=m_cf_\pi\sum_k
\int^\infty_{m_c^2}ds_1
\int^\infty_{m_c^2} ds_2~
\frac{(-1)^ka_k}{\Gamma(k+1)}
(s_1-m_c^2)^k\delta^{(k)}(s_1-s_2) e^{-s_1/M_1^2}e^{-s_2/M_2^2}~.
\label{check}
\ee
Introducing again new variables $s=s_1+s_2$ and $v=s_1/s$
we can use the $\delta$-function to evaluate
the integral over $v$. The result is
\bq
F(M_1^2, M_2^2)= m_cf_\pi\sum_k
\frac{a_k}{2^{k+1}k!}\int_{2m_c^2}^\infty\!\! ds
\left( \frac {d}{ dv} \right)^k\!\!\Bigg[\left(v-\frac{m_c^2}{s}\right)^k
\nonumber
\\
\times
\exp\left(-\frac{svM_2^2+s(1-v)M_1^2}{M_1^2M_2^2}\right)\Bigg]_{v=1/2}~.
\label{form1}
\eq
At $M_1^2=M_2^2=2M^2$ the $v$-dependence of the exponent disappears and
the differentiation of the bracket gives a factor $k!$. We then get
\be
F(M^2)= m_cf_\pi\sum_k \frac{a_k}{2^{k+1}}
\int_{2m_c^2}^\infty ds\,
e^{-\frac{s}{2M^2}}= m_cf_\pi\vp_\pi(1/2)M^2e^{-\frac{m_c^2}{M^2}} ~,
\label{form}
\ee
which is the leading contribution in the sum rule (\ref{fin}).
For arbitrary values of $M_1^2$ and $M_2^2$ a similar
expression is obtained, with the argument of the wave function
and the Borel parameter in eq. (\ref{form}) generalized to
$u_0$ and $M^2$ , respectively, as defined in eq. (\ref{37}).

We now turn to the problem of subtracting the contributions from
excited and continuum states in sum rules.
In the usual approximation based on duality, one identifies the
spectral functions $\rho^{QCD}$ and $\rho^h$ beyond a given
boundary in the $(s_1,s_2)$-plane. Then, the
subtraction effectively amounts to restricting the dispersion integrals
in eq. (\ref{repr}) to the region below this boundary.
Ideally, the result should
not depend on the precise shape of this region.
To be specific, one may take
\be
     s_1^a +s_2^a \leq s_0^a ~,
\label{boundary}
\ee
where $s_0$ plays the role of an effective threshold.
Popular choices of the duality region are triangles
in the $(s_1,s_2)-$ plane corresponding to $a=1$, and
squares corresponding to $a\rightarrow \infty$.
Since the spectral density (\ref{dens}) vanishes everywhere except
at $s_1=s_2$, it is actually irrelevant which form of the boundary
we adopt \footnote{This is
literally true only if the power series defining
the wave function (\ref{sum}) is truncated at some finite order, or
if it converges
rapidly. However, this condition is always met at a sufficiently
high normalization
point where the wave function deviates little from the asymptotic form.
}.

Using duality as outlined above we have to
evaluate the integral in eq. (\ref{check})
with the integration region restricted by eq. (\ref{boundary}).
Changing variables and integrating over $v$ one obtains an
expression similar to eq. (\ref{form1}), but with the upper limit of
integration in $s$ lowered to $2s_0$ and with the addition of surface
terms. The latter disappear for $M_1^2=M_2^2$. Hence,
one is again  led to eq. (\ref{form})  with a simple modification of the
integration limit:
\be
F(M^2)= m_cf_\pi\sum_k \left(\frac{a_k}{2^{k+1}}\right)
\int_{2m_c^2}^{2s_0} ds\,
e^{-\frac{s}{2M^2}}= m_cf_\pi\vp_\pi(1/2)M^2
\left[e^{-\frac{m_c^2}{M^2}}-e^{-\frac{s_0}{M^2}}\right]~.
\label{formh}
\ee
This proves the substitution rule quoted in eq. (\ref{39}).
It is important to note that the proportionality of the  Borel transform
$F(M_1^2, M_2^2)$ given in eq. (\ref{form1}) to the wave function
$\vp_\pi$ at the point $u_0=M_1^2/(M_1^2+M_2^2)$
is generally destroyed by the continuum subtraction. It is
only retained in the symmetric point $M_1^2=M_2^2$ implying $u_0=1/2$.

The above procedure is not possible for higher twist contributions which are
proportional to zero or negative powers of the Borel parameters.
The reason is that the corresponding spectral densities are
not concentrated near the
diagonal $s_1=s_2$. In fact,
the continuum subtraction is rather complicated in these cases.
For further discussion we refer the reader to the second paper of
ref. \cite{BB94}.
Here, we neglect the continuum
subtraction in higher twist terms altogether.
This is justified to a good approximation since the corresponding
spectral densities decrease fast with $s_1$ and $s_2$ as a consequence of
ultraviolet convergence and, hence, the continuum
contribution is expected
to be small anyway.

\appende

\newpage

\begin{table}
\caption{Summary of theoretical estimates. }

\bigskip
\begin{tabular}{lllll}
\hline
\hline
\\
Reference & $\hat{g}$ &
$g_{B^*B\pi}$ & $g_{D^*D\pi}$ &
$\Gamma(D^{*+}\ra D^0\pi^+)$ (keV)\\
\\
\hline
\\
This paper & 0.32 $\pm$ 0.02 & 29 $\pm$ 3 & 12.5 $\pm$ 1.0 & 32 $\pm$ 5\\
\\
This paper$^a$ & -- & 28 $\pm$ 6& 11 $\pm$ 2 & --\\
\\
\cite{ovch}$^a$ &--& 32 $\pm$ 6& -- & --\\
\\
\cite{Cetc94}$^a$ &0.39 $\pm$ 0.16& 20 $\pm$ 4&9 $\pm$ 1& --\\
\\
\cite{Cetc94}$^{a *}$ &0.21 $\pm$ 0.06& 15 $\pm$ 4 &7 $\pm$ 1& 10 $\pm$ 3\\
\\
\cite{NW}$^b$ & 0.7&--& --& --\\
\\
\cite{IW90}$^{b}$ &--& 64&--&--\\
\\
\cite{Yan}$^b$ & 0.75 $\div$ 1.0&--& --& 100 $\div$ 180\\
\\
\cite{CG}$^c$ & 0.6 $\div $ 0.7 & -- & -- & 61 $\div$ 78\\
\\
\cite{Ametal}$^c$ & 0.4 $\div$ 0.7 & -- & -- & --\\
\\
\cite{BardeenHill}$^d$ & 0.3 & -- & -- & --\\
\\
\cite{Eichtetal}$^e$ & --& -- & 16.2&53.4\\
\\
\cite{DoXu}$^f$ & --& -- &19.5 $\pm$ 1.0&76 $\pm$ 7\\
\\
\cite{Kam}$^g$ & --& -- &8.9&16\\
\\
\cite{KN}$^h$ & --& -- &8.2& 13.8\\
\\
Experiment$^i$&--& -- &$<$ 21 & $<$ 89\\
\\
\hline
\hline
\end{tabular}
\\
$^a$ QCD sum rules in external axial field or soft pion limit. \\
$^*$ including perturbative correction to the heavy meson decay constants.  \\
$^b$ Quark model + chiral HQET.  \\
$^c$ Chiral HQET with experimental constraints on $D^*$ decays. \\
$^d$ Extended NJL model + chiral HQET . \\
$^e$ Quark Model + scaling relation. \\
$^f$ Relativistic quark model. \\
$^g$ SU(4) symmetry. \\
$^h$ Reggeon quark-gluon string model.  \\
$^i$ Combination of ACCMOR [33] and CLEO [34] measurements.\\

\end{table}

\end{document}